\documentclass[sigconf,nonacm]{acmart}

\usepackage{enumitem}
\usepackage{balance}
\usepackage{pifont}
\usepackage{stmaryrd}
\usepackage{color}
\usepackage{multirow}
\usepackage{makecell}
\usepackage{float}
\usepackage{subcaption}
\usepackage{url}
\usepackage{algorithm}
\usepackage{algorithmic}
\usepackage{bm}
\usepackage{xspace}
\setlist[itemize]{left=0pt}

\newcommand{\ie}{{\textit{i.e.}},\xspace}
\newcommand{\eg}{{\textit{e.g.}},\xspace}
\newcommand{\etc}{etc.}
\newcommand{\etal}{{\textit{et al.}}}

\renewcommand{\thefootnote}{\fnsymbol{footnote}}

\begin{document}

\title{Zero-Query Adversarial Attack on Black-box Automatic Speech Recognition Systems}

\author{Zheng Fang}
\authornote{Key Laboratory of Aerospace Information Security and Trusted Computing, Ministry of Education, School of Cyber Science and Engineering, Wuhan University}
\affiliation{
  \institution{Wuhan University}
  \city{Wuhan}
  \country{China}
  }
\email{zhengfang618@whu.edu.cn}

\author{Tao Wang}
\authornotemark[1]
\affiliation{
  \institution{Wuhan University}
  \city{Wuhan}
  \country{China}
  }
\email{WTBantoeC@whu.edu.cn}

\author{Lingchen Zhao}
\authornotemark[1]
\authornote{Lingchen Zhao is the corresponding author.}
\affiliation{
  \institution{Wuhan University}
  \city{Wuhan}
  \country{China}
  }
\email{lczhaocs@whu.edu.cn}

\author{Shenyi Zhang}
\authornotemark[1]
\affiliation{
  \institution{Wuhan University}
  \city{Wuhan}
  \country{China}
  }
\email{shenyizhang@whu.edu.cn}

\author{Bowen Li}
\authornotemark[1]
\affiliation{
  \institution{Wuhan University}
  \city{Wuhan}
  \country{China}
}
\email{bowenli0427@whu.edu.cn}

\author{Yunjie Ge}
\authornotemark[1]
\affiliation{%
  \institution{Wuhan University}
  \city{Wuhan}
  \country{China}
}
\email{yunjiege@whu.edu.cn}

\author{Qi Li}
\authornote{Institude for Network Sciences and Cyberspace \& BNRist}
\affiliation{%
  \institution{Tsinghua University}
  \city{Beijing}
  \country{China}
}
\email{qli01@tsinghua.edu.cn}

\author{Chao Shen}
\authornote{School of Cyber Science and Engineering}
\affiliation{%
  \institution{Xi'an Jiaotong University}
  \city{Xi'an}
  \country{China}
}
\email{chaoshen@mail.xjtu.edu.cn}

\author{Qian Wang}
\authornotemark[1]
\affiliation{%
  \institution{Wuhan University}
  \city{Wuhan}
  \country{China}
}
\email{qianwang@whu.edu.cn}

% \thanks{$^{\dagger}$ Lingchen Zhao is the corresponding author.}

\thanks{To appear in the Proceedings of The ACM Conference on Computer and Communications Security (CCS), 2024}

\renewcommand{\shortauthors}{Zheng Fang et al.}
\begin{abstract}
  In recent years, extensive research has been conducted on the vulnerability of ASR systems, revealing that black-box adversarial example attacks pose significant threats to real-world ASR systems.
  However, most existing black-box attacks rely on queries to the target ASRs, which is impractical when queries are not permitted.
  In this paper, we propose ZQ-Attack, a transfer-based adversarial attack on ASR systems in the zero-query black-box setting.
  Through a comprehensive review and categorization of modern ASR technologies, we first meticulously select surrogate ASRs of diverse types to generate adversarial examples.
  Following this, ZQ-Attack initializes the adversarial perturbation with a scaled target command audio, rendering it relatively imperceptible while maintaining effectiveness.
  Subsequently, to achieve high transferability of adversarial perturbations, we propose a sequential ensemble optimization algorithm, which iteratively optimizes the adversarial perturbation on each surrogate model, leveraging collaborative information from other models.
  We conduct extensive experiments to evaluate ZQ-Attack.
  In the over-the-line setting, ZQ-Attack achieves a 100\% success rate of attack (SRoA) with an average signal-to-noise ratio (SNR) of 21.91dB on 4 online speech recognition services, and attains an average SRoA of 100\% and SNR of 19.67dB on 16 open-source ASRs.
  For commercial intelligent voice control devices, ZQ-Attack also achieves a 100\% SRoA with an average SNR of 15.77dB in the over-the-air setting.
\end{abstract}


\maketitle

\section{Introduction}

\begin{table*}[!t]
    \centering
    \caption{Summary of existing audio adversarial attacks.}
    \label{tab:summary}
        \begin{tabular}{c|c|c|c|c|c|c}
            \Xhline{1px}
            Method                                 & Setting                              & Knowledge        & Target ASR \footnotemark[1] & Over-the-line \footnotemark[2]          & Over-the-air           & Queries \footnotemark[3] \\
            \Xhline{1px}
            Carlini \etal~\cite{carlini2018audio}  & White-box                             & Gradient         & $\square$                   & 100\%     & -              & $\sim$1000               \\ \hline
            Commandersong~\cite{217607}            & White-box                             & Gradient         & $\square$                   & 100\%      & $\sim$80\% & $\sim$1000               \\ \hline
            Taori~\etal~\cite{taori2019targeted}   & Black-box                             & Prediction score & $\square$                   & $\sim$40\%       & -              & $\sim$300,000             \\ \hline
            SGEA~\cite{wang2020towards}            & Black-box                             & Prediction score & $\square$                   & 100\%      & -              & $\sim$100,000             \\ \hline
            Devil's Whisper~\cite{chen2020devil}   & Black-box                             & Confidence score & $\triangle \star$           & $\sim$60\% & $\sim$50\% & $\sim$1500               \\ \hline
            Occam~\cite{zheng2021black}            & Black-box                             & Final decision   & $\square \triangle$         & 100\%      & -              & $\sim$10,000              \\ \hline
            KENKU~\cite{291098}                    & Black-box                             & None             & $\triangle \star$           & $\sim$70\%     & $\sim$80\%       & $>$0     \\ \hline
            NI-Occam~\cite{zheng2021black}         & Zero-query black-box \footnotemark[4] & None             & $\star$                     & -              & $\sim$50\%       & 0                        \\ \hline
            TransAudio~\cite{qi2023transaudio}     & Zero-query black-box                  & None             & $\square \triangle$         & $\sim$30\% & -              & 0                        \\ \hline
            ZQ-Attack                              & Zero-query black-box                  & None             & $\square \triangle \star$   & 100\%      & 100\%      & 0                        \\
            \Xhline{1px}
        \end{tabular}
        \\
        \parbox{\textwidth}{
        \footnotesize
        Note that,
            (\romannumeral1) \footnotemark[1]: We use $\square$, $\triangle$, $\star$ to represent open-source ASRs, online speech recognition services, and commercial IVC devices, respectively.
            (\romannumeral2) \footnotemark[2]: In the over-the-line and over-the-air settings, we employ `-' to represent unsuccessful attacks.
            In cases of successful attacks, the success rate of attack is presented.
            (\romannumeral3) \footnotemark[3]:
        ``Queries'' refer to the number of queries made to the target ASR system during the generation process.
            (\romannumeral4) \footnotemark[4]:
            In contrast to the black-box setting, the zero-query black-box setting prohibits any queries to the target ASR system during the generation process.
        }
\end{table*}

Automatic speech recognition (ASR) techniques, which convert spoken language into text, play a crucial role in modern human-computer interactions.
These techniques are now widely employed in online speech recognition services~\cite{azure,whisper} provided by major companies, including Microsoft, OpenAI, etc.
Online repositories also offer a plethora of open-source ASRs for public utilization.
Furthermore, ASRs have also been integrated into commercial intelligent voice control (IVC) devices, such as Apple Siri~\cite{siri} and Amazon Alexa~\cite{alexa}, enabling users to perform various tasks via voice commands.
Unfortunately, similar to other deep neural networks (DNNs) based systems, modern ASRs are vulnerable to adversarial attacks~\cite{szegedy2013intriguing,carlini2018audio,gao2018black}. Attackers can introduce small perturbations into audio samples, causing the target ASR system to produce incorrect results.

\noindent
\textbf{Audio Adversarial Attacks.}
Recently, numerous studies have investigated the practicality and effectiveness of audio adversarial attacks on ASR systems, as summarized in Table~\ref{tab:summary}.
Depending on the accessibility of the attacker to the target ASR systems, audio adversarial attacks can be categorized into white-box and black-box attacks.
The initial works~\cite{carlini2018audio, 217607} employ gradient descent algorithms to generate adversarial audios in the white-box setting, where attackers have full access to the internal information of the target ASR systems.
However, in real-world scenarios, attackers typically lack access to the internal information of the target system, making these white-box attacks often impractical. To achieve attacks in black-box settings, several methods have been proposed to generate adversarial examples based on the limited information acquired through queries to the target ASR system~\cite{taori2019targeted,chen2020devil,zheng2021black,291098}.
However, these methods require huge financial costs and time investments for the queries to generate a single adversarial example, and the large number of highly similar queries makes these attacks easily detectable by the target ASR system, rendering them impractical. 

Therefore, to further enhance the practicality of adversarial attacks, researchers have increasingly turned their attention to transfer-based attacks, which can generate adversarial examples effectively across different target systems, thereby eliminating the need for queries.
However, the performance of existing transfer-based audio adversarial example attacks also exhibits significant limitations. For instance, NI-Occam~\cite{zheng2021black} generates audio adversarial examples on fine-tuned Kaldi models to attack IVC devices, but its attack success rate is quite limited. TransAudio~\cite{qi2023transaudio} optimizes adversarial examples on a surrogate ASR model and can successfully attack black-box ASR systems with similar architectures to the surrogate model. However, it only achieves word-level modifications to the original transcription and has a low success rate on online speech recognition services. These results indicate that current transfer-based attacks still possess limited transferability.

Consequently, we propose the following question: \emph{How to generate audio adversarial examples with high transferability in the challenging zero-query black-box setting?}

\noindent
\textbf{ZQ-Attack.} Our answer to this question is ZQ-Attack, a transfer-based adversarial attack on black-box ASR systems without the need for queries.
Inspired by the ensemble method~\cite{opitz1999popular,liu2016delving,brown2017adversarial}, our core idea is to optimize the adversarial perturbation on diverse surrogate ASRs. 
Ideally, adversarial perturbations optimized concurrently on multiple different surrogate models should contain features that can be captured by these models, making them effective against various ASR systems.

Specifically, ZQ-Attack consists of three stages: surrogate ASRs selection, perturbation initialization, and sequential ensemble optimization.
For the surrogate ASRs selection stage, we conduct an extensive survey of modern ASR systems and observe that different types of ASR systems utilize distinct acoustic models. 
Therefore, we first need to select multiple different types of surrogate ASRs.
Then, to ensure that the generated adversarial perturbations are effective across these surrogate ASRs while maintaining high stealthiness, we propose an adaptive search algorithm that uses scaled target commands to initialize the adversarial perturbation, instead of using zeros or Gaussian noise as in existing methods.
Following the initialization, ZQ-Attack employs a sequential ensemble optimization algorithm to optimize the adversarial perturbations on the sequence of diverse surrogate ASRs collaboratively.
This sequential ensemble algorithm allows for the optimization of adversarial perturbations on each surrogate ASR while concurrently leveraging information from prior surrogate ASRs. Consequently, the generated adversarial perturbations are effective not only for the current surrogate ASR but also for the previous ASRs in the sequence.

We conduct extensive experiments in both over-the-line and over-the-air settings to validate the effectiveness and imperceptibility of our ZQ-Attack.
In the over-the-line setting, ZQ-Attack achieves an average success rate of attack (SRoA) of 100\% and signal-to-noise ratio (SNR) of 21.91dB on four online speech recognition services.
Additionally, ZQ-Attack attains an average SRoA of 100\% and SNR of 19.67dB on 16 open-source ASRs.
In the over-the-air setting, ZQ-Attack achieves an average SRoA of 100\% and an SNR of 15.77dB on two commercial IVC devices.
These results demonstrate that ZQ-Attack can successfully generate audio adversarial examples with high transferability, effectively targeting various ASR systems without requiring any queries.

\noindent
\textbf{Contributions.}
Our contributions are summarized as follows:
\begin{itemize}
    \item
          To the best of our knowledge, we are the first to generate audio adversarial examples with high transferability on ASR systems in the most challenging zero-query black-box setting.
          Our method is effective in both over-the-line and over-the-air settings, with the target ASR systems encompassing online speech recognition services, open-source ASRs, and commercial IVC devices, showcasing remarkable practicality.
    \item
          We introduce ZQ-Attack, a zero-query, transfer-based adversarial attack on black-box ASR systems. This approach optimizes adversarial perturbations using a set of diverse surrogate ASRs simultaneously, thereby enhancing their transferability. Furthermore, we develop an adaptive adversarial perturbation initialization method based on the target command audio to improve the imperceptibility.
 
    \item
          We conduct comprehensive experiments to evaluate the performance of ZQ-Attack on online speech recognition services, open-source ASRs, and commercial IVC devices.
          The experimental results demonstrate the superior performance of our method. ZQ-Attack can successfully attack all the target ASR systems without queries while achieving an average SRoA of 100\%.
\end{itemize}

\section{Background}
\subsection{Automatic Speech Recognition}
\label{section:asr}

\begin{figure}[t!]
  \centering
  \includegraphics[scale=0.45]{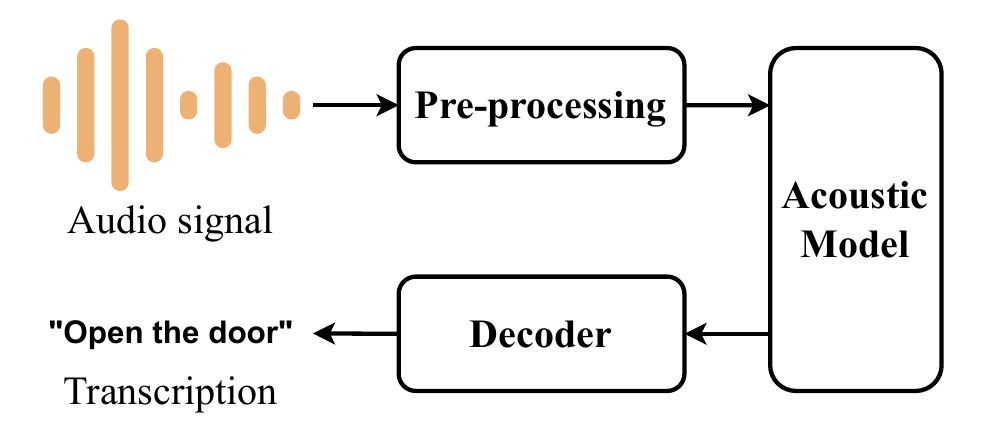}
  \caption{The architecture of a typical ASR system.}
  \label{fig:architecture}
\end{figure}

ASR systems can automatically transcribe input audio into the corresponding transcriptions.
As shown in Figure~\ref{fig:architecture}, a typical ASR system comprises three components: pre-processing, acoustic model, and decoder.

\begin{itemize}
  \item \textit{Pre-processing}.~
        Given an input audio with $n$ sample points, denoted as $x \in \mathbb{Z}^n$, an ASR system first normalizes it to the range of ${[-1,1]} ^ n$ and applies low/high-pass filters to remove frequencies and segments beyond the range of human hearing.
        Then, the ASR system employs time-frequency transformation techniques, such as the short-time Fourier transform (STFT), to convert the time-domain signal $x$ into the frequency domain spectrogram $S \in \mathbb{R} ^ {T \times F}$, where $T$ and $F$ denote the number of frames and frequency bins, respectively.
        In the subsequent steps, traditional ASR systems utilize acoustic feature extraction algorithms, such as mel-frequency cepstral coefficients (MFCC)~\cite{sigurdsson2006mel}, linear predictive coefficient (LPC)~\cite{itakura1975line}, or perceptual linear predictive (PLP)~\cite{hermansky1990perceptual}, to further transform $S$ into meticulously designed acoustic features. In contrast, modern DNN-based ASR systems directly utilize $S$.

  \item \textit{Acoustic model}.~
        The acoustic model is typically a machine learning model that maps the spectrogram or extracted acoustic features to an intermediate representation.
        Traditional ASR systems use hidden Markov models (HMMs) and Gaussian mixture models (GMMs)~\cite{gales2008application,rabiner1989tutorial}, while
        modern ASR systems employ DNNs, such as convolutional neural networks (CNNs) and Transformers~\cite{gulati2020conformer,li2019jasper}.

  \item \textit{Decoder}.~
        The decoder uses the intermediate representation to predict tokens and generate corresponding transcriptions.
        A token is the smallest unit of the transcription, and the set of all possible tokens constitutes a vocabulary $V$.
        The vocabulary $V$ varies across different ASR systems.
        For example, $V = \{a, b, \cdots, z, space\}$ is a simple vocabulary for an ASR system recognizing English.
\end{itemize}

\subsection{Audio Adversarial Attacks}
Audio adversarial attacks aim to manipulate the output of ASR systems using audio adversarial examples constructed by adding imperceptible perturbations to benign carrier audios~\cite{carlini2018audio,217607,zheng2021black,chen2020devil,291098}.
Depending on the objective of the attack, audio adversarial attacks can be categorized into
targeted and untargeted attacks.

Formally, let $f(x): x  \to y $ denote an ASR system that transcribes an audio $x$ into the transcription $y = f(x)$, and let $x^\prime=x+\delta$ denote the adversarial example constructed by adding the adversarial perturbation $\delta$ to the carrier audio $x$.
An untargeted adversarial attack aims to mislead the target ASR system, causing it to produce any result other than the ground truth, represented as $f(x^\prime) \neq y$.
In contrast, a targeted adversarial attack aims to induce the output of the target ASR system into a specific target transcription $t \neq f(x)$, which is formulated as:
\begin{equation}
  f(x^\prime) = t,\quad s.t.~ Dis(x,x') < \epsilon,
\end{equation}
where $Dis(x,x')$ represents the distance between $x$ and $x'$, commonly calculated using the $L_p$ norm, with $p$ usually being 0, 2, or $\infty$.
$\epsilon$ is a hyper-parameter that constrains this distance.
As targeted adversarial attacks can naturally extend to untargeted attacks, the adversarial attacks discussed in the rest of this paper will specifically refer to the targeted ones unless explicitly specified otherwise.

Typically, the generation process of adversarial examples can be formulated as an optimization problem:
\begin{equation}
  \label{equation:traditional_loss}
  \min \limits_{\delta}{\mathcal{L}(x,\delta,t,f)=\mathcal{L}_a(x,\delta,t,f) + c \cdot \mathcal{L}_p(\delta)} ,
\end{equation}
where the adversarial loss $\mathcal{L}_a$ measures the effectiveness of $\delta$ on the target ASR system $f$, and the imperceptibility loss $\mathcal{L}_p$ quantifies the imperceptibility of $\delta$.
The parameter $c$ acts as a weighting factor, balancing the effectiveness and imperceptibility of the attack. Gradient descent is a common method for solving this optimization problem. It can be formulated as:
\begin{equation}
  \label{equation:gradient_descent}
  \delta \leftarrow clip_\epsilon\left(\delta - \alpha \cdot \nabla_\delta\mathcal{L}(x,\delta,t,f)\right),
\end{equation}
where $\alpha$ represents the learning rate and $\nabla_\delta\mathcal{L}(x,\delta,t,f)$ is the gradient of $\mathcal{L}(x,\delta,t,f)$ with respect to $\delta$.
The function $clip_\epsilon$ limits $Dis(x,x')$ to a relatively small range controlled by $\epsilon$.

\section{Threat Model \& Challenges}
\subsection{Threat Model}
\label{section:threat_model}

\begin{figure*}[h!t!]
    \centering
    \includegraphics[scale=1.0]{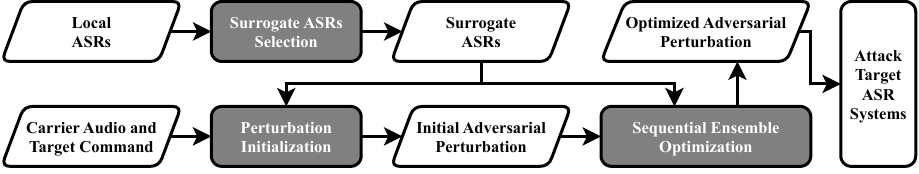}
    \caption{Workflow of ZQ-Attack.
        ZQ-Attack is mainly divided into three stages: surrogate ASRs selection, perturbation initialization, and sequential ensemble optimization.
        }
    \label{The overview}
\end{figure*}

\noindent
\textbf{Goals.~}
The attacker aims to generate audio adversarial examples that can be recognized as the target transcription by the target ASR systems without any queries.
In the over-the-line setting (\ie digital attacks), the target systems are online speech recognition services or open-source ASRs.
The waveform files of the audio adversarial examples are directly used as inputs, inducing the target ASR systems to produce the specified target transcriptions.
In the over-the-air setting (\ie physical attacks), the target ASR systems are commercial IVC devices.
The adversarial examples should be misrecognized as the target transcription by these devices after being transmitted through the air.
Additionally, the audio adversarial examples should be imperceptible, making them difficult for human ears to detect.

\noindent
\textbf{Knowledge \& Capabilities.~}
Prior works on black-box adversarial attacks on ASR systems do not require the attacker to know internal information about the target ASR system, but they still assume that the attacker can interact with the target system. In this paper, we consider a more realistic and challenging scenario where the attacker also cannot query the target ASR system during the generation of adversarial examples.
After generating the adversarial examples, the attacker can execute the attacks by uploading audio files to target ASR systems in the over-the-line setting and positioning a speaker near the target commercial IVC devices in the over-the-air setting.
Additionally, we assume the attacker has the capability to train surrogate ASRs or obtain pre-trained surrogate ASRs from open-source repositories.

\subsection{Challenges}
\label{section:challenges}
To launch audio adversarial attacks in the challenging zero-query black-box setting, an alternative approach is the transfer-based attack. The basic idea of transfer-based attacks is to use a local surrogate model to generate adversarial examples that are then used to attack the target black-box model.
Despite the demonstrated effectiveness of transfer-based attacks in the image domain~\cite{papernot2016transferability,demontis2019adversarial,wu2018understanding,liu2016delving}, achieving similar success in the audio domain remains an unresolved issue.
Abdullah~\etal~\cite{abdullah2021sok} reveal that the transferability of audio adversarial examples among different ASR systems is exceedingly limited, even when these ASRs share the same architecture.
Existing attempts to achieve this goal have only demonstrated limited transferability~\cite{zheng2021black,qi2023transaudio}.
As demonstrated in prior work~\cite{huang2019enhancing,wu2018understanding,ravikumar2022trend}, the core reason for this limitation might be that adversarial examples tend to overfit the architecture and feature representations of the specific surrogate model, resulting in limited transferability to the target models with different architecture.
Unlike image recognition models, ASR systems exhibit increased complexity and greater architectural diversity, leading to greater differences between surrogate ASRs and target ASRs. Hence, generating highly transferable adversarial examples in the audio domain is more challenging.

\section{ZQ-Attack}

\subsection{Problem Formulation}
ZQ-Attack aims to generate transferable audio adversarial examples in the zero-query black-box scenario.
Formally, given a carrier audio $x$ and a target command $t$, ZQ-Attack optimizes the adversarial perturbation $\delta$ to enhance its effectiveness on various black-box target ASR systems. This optimization problem can be formulated as follows:
\begin{equation}
    \max_\delta \mathop{\mathbb{P}}_{f\in\mathcal{F}}~\left(f(x+\delta)=t\right),
\end{equation}
where $\mathbb{P}$ represents the probability, and $\mathcal{F}$ denotes the set of all black-box target ASR systems.
However, since the attacker lacks internal information about the target ASR system and cannot query it, directly solving this optimization problem is challenging.

An intuitive way to solve this problem is leveraging surrogate ASRs. However, optimizing the adversarial perturbation on a single surrogate ASR may result in overfitting to that specific model. Therefore, ZQ-Attack optimizes adversarial perturbations on multiple surrogate ASRs.
We denote the set of surrogate ASRs as $\mathbb{F}$.
Then, the optimization problem becomes as follows:
\begin{equation}
    \min_\delta  \mathcal{L}_{all}(x,\delta,t,\mathbb{F}) \quad \text{s.t.}~ Dis(x,x') < \epsilon,
\end{equation}
where $\mathcal{L}_{all}$ denotes the loss on all surrogate ASRs, and the imposed constraint ensures that the optimized adversarial perturbation attains a specified level of imperceptibility.

\subsection{Attack Overview}

\begin{figure}[!t]
    \centering
    \includegraphics[scale=0.8]{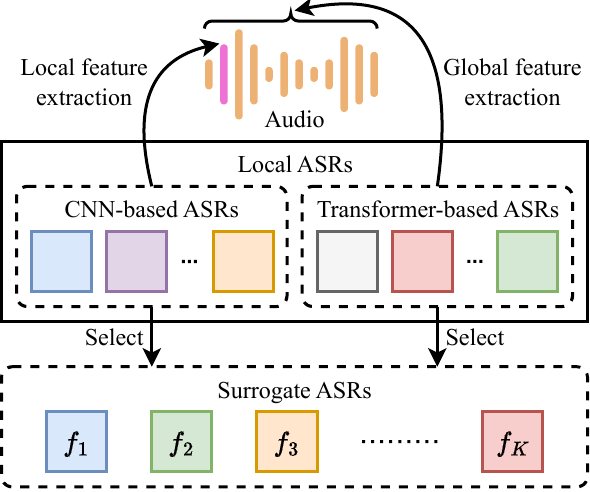}
    \caption{Illustration of surrogate ASRs selection.}
    \label{fig:SAS}
\end{figure}

The core idea of ZQ-Attack is to collaboratively optimize the adversarial perturbation on diverse types of surrogate ASRs.
Specifically, ZQ-Attack consists of three stages: surrogate ASRs selection, perturbation initialization, and sequential ensemble optimization.
The workflow is illustrated in Figure~\ref{The overview}.

\noindent
\textbf{Surrogate ASRs Selection.~}
We investigate modern ASR systems and categorize them into two main types: CNN-based and Transformer-based. While CNNs are more adept at capturing local features, Transformers excel at capturing global contexts.
Therefore, an intuitive approach is to select surrogate ASRs that include both CNN-based and Transformer-based architectures, ensuring that the adversarial perturbations optimized on these surrogate ASRs concurrently possess both local and global features of the target command.

\noindent
\textbf{Perturbation Initialization.~} Given a target command $t$, we employ Text-to-Speech (TTS) techniques to generate a corresponding target command audio $x_t$.
Subsequently, we initialize the adversarial perturbation with a scaled $x_t$ and superimpose it onto the carrier audio, ensuring that the constructed adversarial example is effective on all surrogate ASRs.
To further enhance the imperceptibility of the initial adversarial perturbation, we employ an adaptive search algorithm to minimize the scaling factor.

\noindent
\textbf{Sequential Ensemble Optimization.~}
Following the perturbation initialization stage, ZQ-Attack employs a sequential ensemble optimization algorithm to collaboratively optimize the adversarial perturbation on the ordered set of surrogate ASRs.
This algorithm consists of an inner loop and an outer loop.
In each iteration, the ordered set of surrogate ASRs is randomly shuffled in the outer loop. Then, the sequential ensemble optimization takes place within the inner loop. For each surrogate ASR, this algorithm integrates collaborative information from all preceding surrogate ASRs in the ordered set, facilitating collaborative optimization.
Additionally, we design a novel loss function for the optimization process to enhance the transferability and imperceptibility
of the perturbation.
Following the inner loop, the algorithm updates the perturbation and validates its effectiveness on all surrogate ASRs.

\subsection{Surrogate ASRs Selection}
\label{subsection:The Selection of Surrogate ASRs}
The selection of surrogate ASRs is the foundation that ensures the effectiveness of ZQ-Attack.
As described in Section~\ref{section:challenges}, the architecture of ASR systems exhibits considerable diversity.
Therefore, randomly selecting surrogate ASRs may fail to cover the mainstream modern ASR systems.
Hence, in this subsection, we initially provide a summary and categorization of modern ASR systems, followed by the details of surrogate ASRs selection. The illustration of this stage is shown in Figure~\ref{fig:SAS}.

\noindent
\textbf{Summary and Categorization of Modern ASR Systems.~}
Modern ASR systems typically convert audio into the spectrogram without employing additional acoustic feature extraction algorithms~\cite{kriman2020quartznet,li2019jasper,han2020contextnet,gulati2020conformer,whisper,majumdar2021citrinet,amodei2016deep,hannun2014deep,zhang2020transformer}. Hence, the primary differences among these ASR systems reside in the internal acoustic model. 
We summarize and categorize modern ASR systems into the following two categories from the perspective of the acoustic model:
\begin{itemize}
    \item \textit{CNN-based.~}
    CNNs have found widespread application in the field of computer vision and have recently exhibited notable progress in ASR as well~\cite{li2019jasper,kriman2020quartznet,han2020contextnet,majumdar2021citrinet,abdel2014convolutional,abdeljaber2017real,DBLP:conf/interspeech/SchneiderBCA19}.
    The key advantages of CNNs include their model's low complexity and high computational efficiency. Additionally, CNNS are adept at extracting local features within the spectrogram.

    \item \textit{Transformer-based.~}
    These ASR systems employ Transformers as their acoustic models. Given that audio is temporal data, using recurrent neural networks (RNNs) as acoustic models is an intuitive choice~\cite{amodei2016deep,hannun2014deep}.
    While RNNs can capture short-term dependencies, attention mechanisms~\cite{vaswani2017attention} enable non-contiguous frames to attend each other, allowing Transformers to capture long-term dependencies. This results in better speech recognition performances than RNNs~\cite{zeyer2019comparison,karita2019comparative,li2020comparison}.
    Consequently, Transformers are progressively supplanting RNNs in modern ASRs~\cite{dong2018speech,zhang2020transformer,karita2019improving,DBLP:conf/interspeech/Higuchi0COK20,gulati2020conformer,baevski2020wav2vec,hsu2021hubert,radford2023robust}.
\end{itemize}

\noindent
\textbf{Selection of Surrogate ASRs.~}
The local ASRs for selection can be obtained from online sources or trained by the attackers themselves.
While utilizing a single surrogate ASR can lead to the overfitting of $\delta$ to that surrogate ASR, rendering $\delta$ ineffective on target ASR systems, using too many surrogate ASRs can also lead to high computation costs. The diversity in architectures of ASRs may result in significant disparities between the surrogate ASRs and the target ASR systems, making it challenging to generate transferable adversarial examples.
Therefore, we need to select surrogate ASRs encompassing modern ASR systems of different types.

According to our categorization, modern ASR systems can be mainly categorized into CNN-based and Transformer-based ASR systems.
The CNNs excel in extracting local features but exhibit a comparatively weaker capability in capturing dynamic global contexts.
Conversely, Transformers are proficient in effectively capturing global information but demonstrate a diminished ability to extract local features.
To integrate the advantages of both CNN-based and Transformer-based ASR systems, the selected surrogate ASRs should encompass representatives from both categories.
This ensures that the optimized adversarial perturbations can possess both locally and globally salient features, thereby enhancing their transferability to the target ASR systems.
Furthermore, CNNs and Transformers represent prevalent architectures of acoustic models adopted in modern ASR systems.
Therefore, we incorporate both CNN-based and Transformer-based ASRs into the surrogate ASRs to effectively cover a broad range of real-world ASR systems.
Upon selecting $K$ surrogate ASRs from the local ASRs, we construct the set of these surrogate ASRs, denoted as $\mathbb{F}=[f_j]_{j=1}^K$, where $f_j$ represents the $j$-th surrogate ASR in $\mathbb{F}$.
Since the subsequent sequential ensemble optimization algorithm optimizes the adversarial perturbation on these surrogate ASRs in a sequential manner, $\mathbb{F}$ is an ordered set.
It is worth noting that the surrogate ASRs are scalable.
The quantity of surrogate ASRs can be adjusted flexibly depending on the computational resources of the attackers.

\begin{algorithm}[!t]
    \caption{Adaptive Search (AdaSearch)}
    \label{alg:adaptive_search}
    \renewcommand{\algorithmicrequire}{\textbf{Input:}}
    \renewcommand{\algorithmicensure}{\textbf{Output:}}
    \begin{algorithmic}[1]
        \REQUIRE Carrier audio $x$, Target command $t$, Target command audio $x_t$, Ordered set of $K$ surrogate ASRs $\mathbb{F}$, Search stride $s$
        \ENSURE Initial adversarial perturbation $\delta$
        \STATE $l_x \leftarrow len(x),~l_t \leftarrow len(x_t)$
        \STATE $\mu = +\infty,~\delta \leftarrow 0$
        \FOR{$i \leftarrow 0$ \TO\ $l_x-l_t$}
            \STATE $\mu_i \leftarrow 0$
            \STATE $\delta_i \leftarrow [\underbrace{0,\cdots,0}_{i},~x_t~,\underbrace{0,\cdots,0}_{l_x-l_t-i}]$
            \WHILE {$\mu_i \le \mu$}
            \IF{$\forall f \in \mathbb{F},~f(x+\mu_i \cdot \delta_i) = t$}
                \STATE $\mu = \mu_i,~\delta = \mu \cdot \delta_i$
                \STATE \textbf{break}
            \ENDIF
            \STATE$\mu_i \leftarrow \mu_i + s$
            \ENDWHILE
        \ENDFOR
        \RETURN $\delta$
    \end{algorithmic}
\end{algorithm}

\subsection{Perturbation Initialization}
The initialization of the adversarial perturbation $\delta$ may significantly impact the performance of the attack.
Initializing $\delta$ from a point far from the region of the target command in the feature space can lead to a time-consuming and uncertain optimization process, and the high dimensionality of audio data further complicates the optimization.
In contrast, using the target command audio directly as the initial $\delta$ may result in poor imperceptibility.
To obtain an effective and relatively imperceptible initialized adversarial perturbation, we propose an adaptive search algorithm to initialize $\delta$ with a scaled target command audio, as presented in Alg.~\ref{alg:adaptive_search}.
This algorithm aims to minimize the scaling factor while maintaining the effectiveness of the initialized $\delta$ on all surrogate ASRs.

Specifically, we first choose an audio $x$ as the carrier audio to construct the adversarial example $x'=x + \delta$. Following previous works~\cite{zheng2021black,217607}, we opt for songs as the carrier audio. For a given target command $t$, we utilize TTS techniques to generate a corresponding target command audio $x_t=\mathcal{T}(t)$, where $\mathcal{T}(\cdot)$ denotes the TTS process.
To alleviate the impact of varying volume levels during the perturbation initialization stage, we normalize the values of sample points in both $x$ and $x_t$ to the range of $[-0.5,0.5]$.
Subsequently, the adaptive search algorithm searches for the smallest value of scaling factor $\mu$, and $\delta$ is initialized using the scaled target command audio $\mu \cdot x_t$, ensuring that the corresponding initial adversarial example is recognized by all surrogate ASRs as the target command.

\begin{figure}[t!]
    \centering
    \includegraphics[scale=0.41]{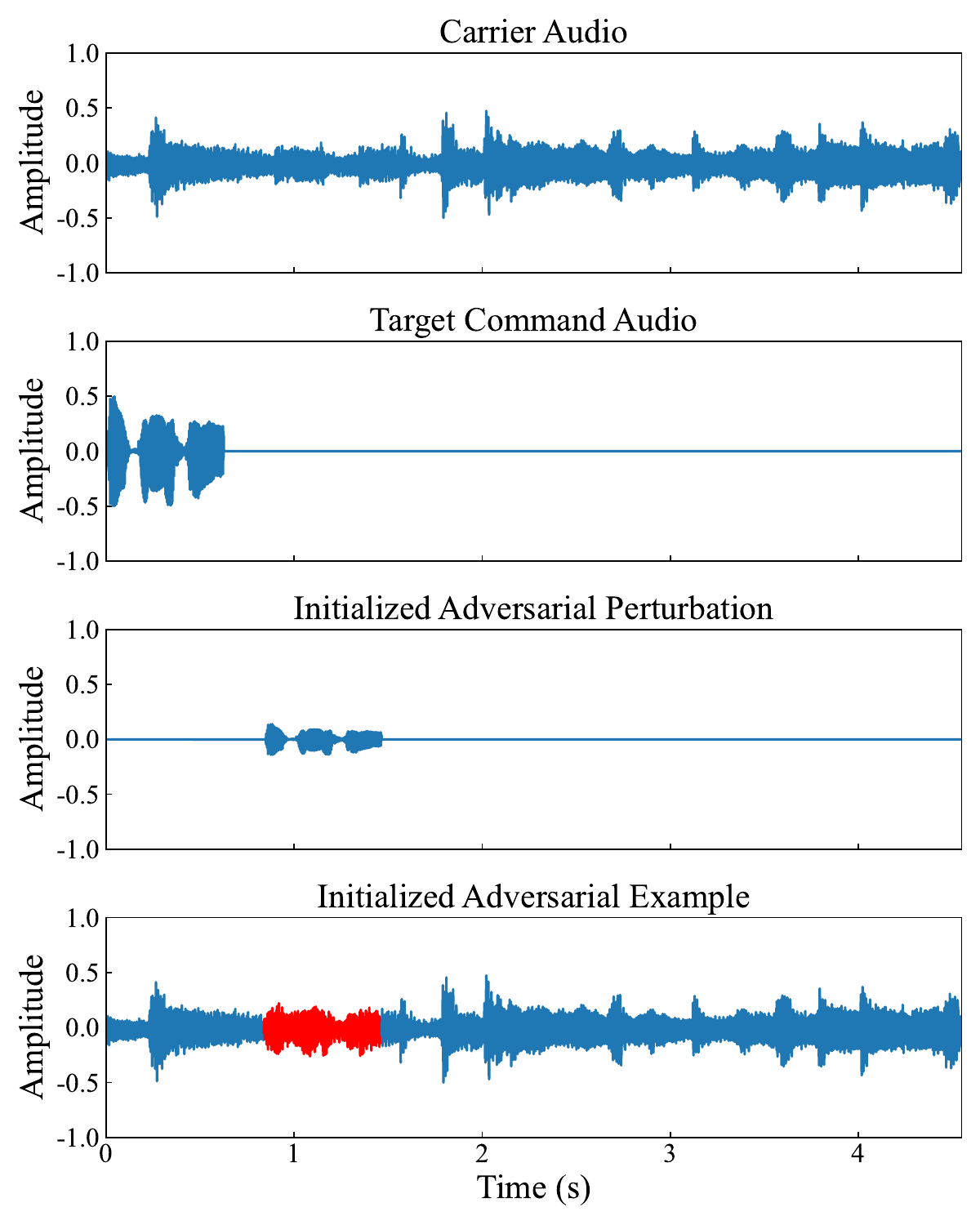}
    \caption{An example of the perturbation initialization. The adversarial perturbation is initialized using a scaled target command audio. The region of the added initialized adversarial perturbation is highlighted in red.}
    \label{imperceptibility}
\end{figure}

Since the length of $x_t$ is typically shorter than that of $x$, it is necessary to pad both sides of the scaled $x_t$ with zeros to initialize the perturbation. We use $l_x$ and $l_t$ to denote the length of $x$ and $x_t$, respectively. The lengths of the padding on each side are indeterminate, provided their sum equals $l_x - l_t$. The adaptive search algorithm searches for the optimal padding lengths on each side to minimize the scaling factor as much as possible.
An example of this initialization method is depicted in Figure~\ref{imperceptibility}.
It can be seen that the adaptive search algorithm finds the padding lengths and a relatively small scaling factor.

In summary, the adaptive search algorithm initializes $\delta$ by pushing the corresponding adversarial example toward the decision boundary of all surrogate ASRs, thereby circumventing the time-consuming and uncertain initial search process.
The initialized adversarial perturbation can be regarded as a coarse-grained optimized perturbation, serving as a basis for subsequent fine-grained optimization in the sequential ensemble optimization stage.

\begin{figure*}[!t]
    \centering
    \includegraphics[scale=0.9]{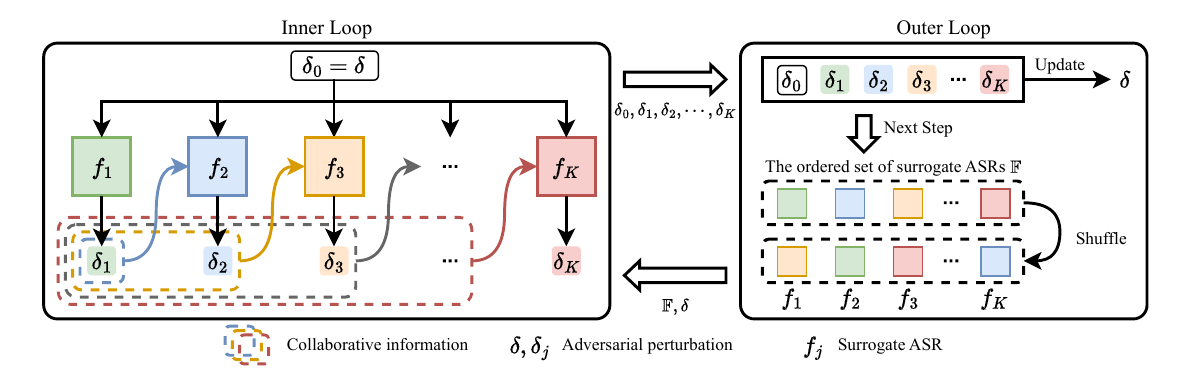}
    \caption{Illustration of sequential ensemble optimization. 
    }
    \label{fig:SEO}
\end{figure*}

\subsection{Sequential Ensemble Optimization}
\label{section:cross_model_optimazation}

After initializing the adversarial perturbation, ZQ-Attack performs fine-grained optimization of the adversarial perturbation on surrogate ASRs.
Unlike the target black-box ASR systems, where the attacker lacks knowledge of their internal architectures and parameters, the white-box surrogate ASRs provide full control to the attacker.
Hence, the attacker can optimize $\delta$ using any information acquired through these surrogate ASRs.

A straightforward approach to optimizing $\delta$ on diverse surrogate ASRs is to use the weighted average of the gradients from each one. For the $j$-th surrogate ASR $f_j \in \mathbb{F}$, the gradient is calculated as $\nabla_{\delta}\mathcal{L}(x,\delta,t,f_j)$, where $\mathcal{L}$ represents the loss of $\delta$ on a single surrogate ASR $f_j$, typically similar to Eq.~\eqref{equation:traditional_loss}.
However, this method essentially treats each surrogate ASR independently when optimizing $\delta$. Each surrogate ASR does not interact with the others and, therefore, cannot leverage the optimization information provided by others.
Moreover, this method focuses on optimizing $\delta$ in the direction most effective for a single surrogate ASR, without considering the effectiveness of $\delta$ on other surrogate ASRs.

To facilitate collaboration among these surrogate ASRs, we propose a sequential ensemble optimization algorithm, as presented in Figure~\ref{fig:SEO}.
This algorithm iteratively optimizes the adversarial perturbation on the ordered set of surrogate ASRs $\mathbb{F}$.
For each surrogate ASR, this algorithm leverages the collaborative information from the preceding surrogate ASRs in the ordered set to optimize $\delta$.
In other words, the optimization process considers not only the efficacy of $\delta$ on the current surrogate ASR but also its efficacy on the preceding surrogate ASRs.
Additionally, instead of directly using $\mathcal{L}$ in Eq.~\eqref{equation:traditional_loss}, we carefully design a novel loss function comprising three loss terms. The detailed design of the loss function is presented in Section~\ref{subsection: loss}.

Specifically, the sequential ensemble optimization algorithm comprises an outer loop and an inner loop.
At each step, the algorithm first randomly shuffles $\mathbb{F}$ in the outer loop.
Then, the optimization of $\delta$ on $\mathbb{F}$ takes place in the inner loop.
Following the completion of the inner loop, $\delta$ is updated in the outer loop.
As ZQ-Attack abstains from interacting with the target ASR systems via queries, it relies on surrogate ASRs to validate the efficacy of the generated adversarial example $x'$. The set of valid adversarial examples, denoted as $X'$, is initialized as $\varnothing$ at the beginning of the algorithm. At the end of each step, $x'$ will be added to $X'$ if it can successfully attack all surrogate ASRs.

\noindent
\textbf{Inner Loop.}
For clarity, we use $\delta_{j}$ to denote the adversarial perturbation optimized on the previous $j$ surrogate ASR(s) in $\mathbb{F}$, and $\delta_{0}$ is equivalent to $\delta$.
For the $j$-th surrogate ASR $f_j$ in $\mathbb{F}$, the input perturbations include $\delta_0$ and the optimized adversarial perturbations obtained from the preceding $j-1$ ASR(s).
Formally, the input adversarial perturbations for $f_j$, denoted as $\Delta_j$, can be represented as $\Delta_j=[\delta_0,\delta_{1},\cdots,\delta_{j-1}]$.
We first add the adversarial perturbations $\Delta_j$ to the carrier audio $x$ to construct the adversarial examples.
To ensure that the adversarial examples have been pushed towards the decision boundary, we additionally add randomly sampled Gaussian noise $\sigma$ on each adversarial example.
Following the forwarding of these adversarial examples to $f_j$ and the subsequent gradient calculation, the perturbation is updated as:
\begin{equation}
    \label{equation:perturbation_update}
    \delta_{j} = \delta_0 - \alpha \cdot \frac{1}{j} \sum_{\delta^\prime \in \Delta_j} \nabla_{\delta^\prime}\mathcal{L}(x,{\delta^\prime}+\sigma,t,f_j),
\end{equation}
where $\alpha$ represents the learning rate. The loss $\mathcal{L}$ is detailed in Section~\ref{subsection: loss}.

To avoid the updated $\delta_j$ being perceptible enough to the human ear, we use a clipping algorithm to restrict the updated $\delta_j$ within a limited range.
Instead of using the $L_p$ norm-based clipping algorithm employed in previous work~\cite{zheng2021black,chen2020devil}, we utilize an adaptive clipping algorithm based on psychoacoustics~\cite{gelfand2013hearing}.
To be specific, this algorithm is grounded in temporal masking, a phenomenon where the presence of louder components can influence the perception of quieter components when two sounds with different loudness levels are heard by the human ear.
This algorithm constrains $\delta$ within a range proportional to the carrier audio, permitting larger perturbations in louder segments of the carrier while maintaining smaller perturbations in quieter areas, thereby reducing the perceptibility of the perturbation. This clipping algorithm can be represented as follows:
\begin{equation}
    \label{equation:clip}
    clip_\epsilon(\delta, x) = \text{max}(\text{min}(\delta, \epsilon \cdot |x|),\ - \epsilon \cdot |x|).
\end{equation}

It is noteworthy that the clipping is performed element-wise in the perturbation. After applying the adaptive clipping algorithm, the adversarial perturbation will be bounded as $\delta_{j} = clip_\epsilon(\delta_{j},x)$.

\noindent
\textbf{Outer Loop.}
Each step of the sequential ensemble optimization algorithm begins by randomly shuffling $\mathbb{F}$.
Subsequently, the optimization is performed in the inner loop.
Following the inner loop, we obtain $K$ perturbations $[\delta_{1}, \delta_{2}, \cdots, \delta_{K}]$ optimized on the surrogate ASRs, and $\delta$ is updated as follows:
\begin{equation}
    \label{equ: update}
    \delta = \eta \cdot \frac{1}{K}\sum_{j=1}^K \delta_{j} + (1-\eta) \cdot \delta_0,
\end{equation}
where $\eta$ denotes a balancing factor determining the extent of the update.
$\eta=0$ implies no update, while $\eta=1$ signifies a complete update to the optimized adversarial perturbation, disregarding the former one.
This update method uses the adversarial perturbation $\delta_0$ as historical information, while the momentum method~\cite{dong2018boosting} uses the gradient as historical information.
At the end of each step, $x'=x+\delta$ is added to $X'$ if all surrogate ASRs transcribe it into the target command.
% it successfully attacks all surrogate ASRs.

The comprehensive details of the sequential ensemble optimization algorithm are presented in Alg.~\ref{alg:ZQ-Attack}.

\begin{algorithm}[!t]
    \caption{ZQ-Attack}
    \label{alg:ZQ-Attack}
    \renewcommand{\algorithmicrequire}{\textbf{Input:}}
    \renewcommand{\algorithmicensure}{\textbf{Output:}}
    \begin{algorithmic}[1]
        \REQUIRE Carrier audio $x$, Target command $t$, Ordered set of $K$ surrogate ASRs $\mathbb{F}$, Max step $N$, Search stride $s$
        \ENSURE The set of valid adversarial examples $X'$
        \STATE $x_t \leftarrow \mathcal{T}(t)$
        \STATE $\delta \leftarrow \text{AdaSearch}(x,t,x_t,\mathbb{F},s)$
        \STATE $X' \leftarrow \varnothing$
        \STATE \textbf{\# Outer Loop}
        \FOR{$i \leftarrow 1$ \TO\ $N$}
            \STATE Randomly shuffle $\mathbb{F}$
            \STATE $\delta_0 \leftarrow \delta$
            \STATE \textbf{\# Inner Loop}
            \FOR{$j \leftarrow 1$ \TO\ $K$}
                \STATE $f_j \leftarrow $ the $j$-th ASR in $\mathbb{F}$
                \STATE $\Delta_j \leftarrow [\delta_0, \delta_1, \delta_2, \cdots, \delta_{j-1}]$
                % \STATE $X_j \leftarrow x + \Delta_j$
                \STATE Sample a Gaussian noise $\sigma$ 
                \STATE Compute $\nabla_{\delta^\prime}\mathcal{L}(x,{\delta^\prime}+\sigma,t,f_j)$ for each $\delta^\prime$ in $\Delta_j$
                \STATE Update $\delta_{j}$ using Eq.~\eqref{equation:perturbation_update}
                \STATE Clip $\delta_{j}$ using Eq.~\eqref{equation:clip}
            \ENDFOR
            \STATE Update $\delta$ using Eq.~\eqref{equ: update}
            \STATE $x'=x+\delta$
            \IF{$\forall f \in \mathbb{F},~f(x') = t$}
                \STATE $X' \leftarrow X' \cup x'$
            \ENDIF
        \ENDFOR
        \RETURN $X'$
    \end{algorithmic}
\end{algorithm}

\subsection{Loss Design}
\label{subsection: loss}
We design a novel loss $\mathcal{L}$ for optimizing the adversarial perturbations. It comprises three terms: the adversarial loss $\mathcal{L}_a$, the imperceptibility loss $\mathcal{L}_p$, and the acoustic feature loss $\mathcal{L}_f$.
The loss $\mathcal{L}$ can be written as follows:
\begin{equation}
    \label{equation:ZQ-Attack_loss}
    \mathcal{L}(x,\delta,t,f) = \mathcal{L}_a(x,\delta,t,f) + c_1 \cdot \mathcal{L}_p(x,\delta) + c_2 \cdot \mathcal{L}_f(x,\delta,t),
\end{equation}
where $c_1$ and $c_2$ serve as weighting factors to balance the relative importance of different loss terms, ensuring a trade-off between the effectiveness and imperceptibility of $\delta$.

\noindent
\textbf{Adversarial Loss.~}
The adversarial loss $\mathcal{L}_{a}$ measures the effectiveness of $\delta$ on a surrogate ASR, 
evaluating how accurately a specific surrogate ASR transcribes the adversarial example to match the target command.
% Therefore, $\mathcal{L}_{a}$ is related to the target ASR.
The calculation process of $\mathcal{L}_{a}$ begins by inputting the constructed adversarial example into the surrogate ASR to obtain the output probability. Then, $\mathcal{L}_{a}$ is calculated according to the output probability and $t$.
As different ASR systems may utilize different loss functions for training, such as connectionist temporal classification (CTC)~\cite{graves2006connectionist} and Transducer~\cite{graves2012sequence, graves2013speech}, the calculation method of $\mathcal{L}_{a}$ varies depending on the specific surrogate ASR.
For instance, we utilize CTC loss to compute $\mathcal{L}_{a}$ for a surrogate ASR with a CTC architecture (\eg Citrinet~\cite{majumdar2021citrinet}).

\noindent
\textbf{Imperceptibility Loss.~}
The imperceptibility loss $\mathcal{L}_{p}$ aims to minimize the detectability of the adversarial perturbation by human ears.
Previous works~\cite{chen2020devil,zheng2021black,291098} commonly utilize the $L_p$ norm of the adversarial perturbation as the imperceptibility loss.
However, Duan~\etal~\cite{duanPerception} demonstrated that the $L_p$ norm shows a limited correlation with human perception, and the $L_2$ norm exhibits a relatively high correlation with human perception among the $L_p$ norm.
Similar to the adaptive clipping algorithm employed when clipping the adversarial perturbation, as shown in Eq.~\eqref{equation:clip}, we design a new imperceptibility loss function $\mathcal{L}_{p}$ to calculate the $L_2$ norm of the ratio of the adversarial perturbation to the carrier audio.
Formally, it can be represented as:
\begin{equation}
    \mathcal{L}_{p}(x,\delta)= {\left\Vert \frac{\delta}{x}\right\Vert}_2.
\end{equation}

\noindent
\textbf{Acoustic Feature Loss.~}
As mentioned in Section~\ref{section:asr}, traditional ASR systems employ feature extraction algorithms to obtain the acoustic features of spectrograms as a preprocessing procedure, while modern ASR systems typically utilize the spectrograms directly.
Despite the intricacy of acoustic feature extraction algorithms, which require specialized knowledge, the performance of traditional ASR systems demonstrates that these algorithms can extract high-quality features.
Prior work has also demonstrated the effectiveness of incorporating acoustic features into the optimization process of audio adversarial examples~\cite{291098}.
Hence, we utilize the acoustic feature loss based on the widely adopted acoustic feature extraction algorithm, MFCC~\cite{sigurdsson2006mel}, to further enhance the effectiveness of the adversarial perturbation.
Specifically, we first extract the acoustic features of the target command audio and the constructed adversarial example.
We denote the acoustic feature of the target command audio and the constructed adversarial example as $M_t$ and $M_{x^\prime}$, respectively.
Then, the acoustic feature loss $\mathcal{L}_{f}$ can be calculated as:
\begin{equation}
    \mathcal{L}_{f}(x,\delta,t)= {\Vert M_{x^\prime}-M_t \Vert}_2.
\end{equation}

\renewcommand{\thefootnote}{\fnsymbol{footnote}}
\begin{table*}[!t]
    \caption{Summary of target ASR systems in the experiments.}
    \label{tab:models}
    \centering
    \resizebox{\linewidth}{!}{
            \begin{tabular}{c|c|c|c}
                \Xhline{1px}
                ASR                                             & Type                              & Acoustic Model    & Word Error Rate on LibriSpeech test-clean/test-other (\%)                               \\ \Xhline{1px}
                Jasper~\cite{li2019jasper}                      & Open-source ASR                   & CNN               & 3.9/12.0                                                                    \\ \hline
                QuartzNet~\cite{kriman2020quartznet}            & Open-source ASR                   & CNN               & 3.8/10.4                                                                    \\ \hline
                Citrinet~\cite{majumdar2021citrinet}            & Open-source ASR                   & CNN               & 4.4/10.7~(S) 3.7/8.9~(M) 3.6/7.9~(L)                                           \\ \hline
                ContextNet~\cite{han2020contextnet}             & Open-source ASR                   & CNN               & 3.3/8.0~(S) 2.2/5.0~(M) 1.9/3.9~(L)                                            \\ \hline
                Conformer-CTC~\cite{gulati2020conformer}        & Open-source ASR                   & Transformer       & 3.7/8.1~(S) 2.6/5.9~(M) 2.1/4.5~(L) 2.0/3.7~(XL)                                \\ \hline
                Conformer-Transducer~\cite{gulati2020conformer} & Open-source ASR                   & Transformer       & 2.9/6.6~(S) 2.1/4.7~(M) 1.7/3.7~(L) 1.6/3.0~(XL)                                \\ \hline
                Whisper~\cite{radford2023robust}                & Open-source ASR                   & Transformer       & 5.0/12.4~(base) 3.4/7.6~(small) 2.9/5.9~(medium) 2.7/5.6~(large)                \\ \hline
                Microsoft Azure~\cite{azure}                    & Online speech recognition service & -\footnotemark[7] & -                                                                           \\ \hline
                Tencent Cloud~\cite{tencent}                    & Online speech recognition service & -                 & -                                                                           \\ \hline
                Alibaba Cloud~\cite{ali}                        & Online speech recognition service & -                 & -                                                                           \\ \hline
                OpenAI Whisper~\cite{whisper} \footnotemark[6]  & Online speech recognition service & Transformer       & 2.7/5.2                                                                     \\ \hline
                Apple Siri~\cite{siri}                         & Commercial IVC device             & -                 & -                                                                           \\ \hline
                Amazon Alexa~\cite{alexa}                       & Commercial IVC device             & -                 & -                                                                           \\ \Xhline{1px}
            \end{tabular}
            }
            \\
            \parbox{\textwidth}{
            \footnotesize
            Note that,
                (\romannumeral1) \footnotemark[7]: Most online speech recognition services and commercial IVC devices do not reveal their implementation of the underlying ASR systems.
                Hence, we lack knowledge about their acoustic models and related information.
                We use ``-'' to represent unknown information.
                (\romannumeral2) \footnotemark[6]: The ASR system employed by the OpenAI API is the open-source Whisper large-v2.
                Therefore, we have access to information regarding the acoustic model and its recognition performance on LibriSpeech.
                Despite the open-source nature of Whisper large-v2, we treat it as a black-box ASR system during attacks.
            }
\end{table*}

\section{Experiments}

\subsection{Experiment Setup}
In this section, we evaluate the performance of ZQ-Attack in both over-the-line and over-the-air settings.
In the over-the-line setting, the audio adversarial examples are transmitted directly to the target ASR systems as waveform audio files.
In the over-the-air setting, we utilize a speaker positioned near the target devices (\eg 10 cm) to play the audio adversarial examples in a quiet office environment (\eg 35dB).

\noindent
\textbf{Target ASR Systems.}
To fully demonstrate the effectiveness of ZQ-Attack, we conduct extensive experiments on various online speech recognition services, commercial IVC devices, and open-source ASRs.
The details are presented as follows.
\begin{itemize}
    \item \textit{Online speech recognition services.}
          In the over-the-line setting, the target online speech recognition services include Microsoft Azure Speech Service~\cite{azure}, Tencent Cloud Automatic Speech Recognition~\cite{tencent}, Alibaba Cloud intelligent speech interaction~\cite{ali}, and OpenAI Whisper~\cite{whisper}.
    \item \textit{Commercial IVC devices.}
          In the over-the-air setting, we select Apple Siri~\cite{siri}, and Amazon Alexa~\cite{alexa} as the target IVC devices.
    \item \textit{Open-source ASRs.}
          Among the numerous open-source ASRs, we select some of the most representative and advanced ASRs as our targets in the over-the-line setting.
          As prior research indicates that the transferability of audio adversarial examples can be limited even among instances of the same ASR~\cite{abdullah2021sok},
          we select open-source ASRs with the same architectures but varying scales, as well as those with distinct architectures. 
          Specifically, the target ASRs include: Jasper~\cite{li2019jasper}, QuartzNet~\cite{kriman2020quartznet}, ContextNet (M/L)~\cite{han2020contextnet}, Citrinet (M/L)~\cite{majumdar2021citrinet}, Conformer-CTC (M/L/XL)~\cite{gulati2020conformer}, Conformer-Transducer (M/L/XL)~\cite{gulati2020conformer}, and Whisper (base, small, medium, large)~\cite{radford2023robust}.
\end{itemize}

We summarize and categorize these target systems in Table~\ref{tab:models}.

\noindent
\textbf{Surrogate ASRs.}
The surrogate ASRs include Conformer-CTC (S)~\cite{gulati2020conformer}, Conformer-Transducer (S)~\cite{gulati2020conformer}, ContextNet (S)~\cite{han2020contextnet}, and Citrinet (S)~\cite{majumdar2021citrinet}.
ContextNet and Citrinet employ CNNs as the acoustic model, while the acoustic models of Conformer-CTC and Conformer-Transducer are based on Transformers.
The checkpoints for these surrogate ASRs are obtained from Nvidia NeMo~\cite{Nemo}.

\noindent
\textbf{Target Commands and Carrier Audios.}
We choose 10 commonly used instructions as the target commands in the experiments: \textit{call my wife}, \textit{make it warmer}, \textit{navigate to my home}, \textit{open the door}, \textit{open the website}, \textit{play music}, \textit{send a text}, \textit{take a picture}, \textit{turn off the light}, and \textit{turn on airplane mode}.
We select five songs used in Commandersong~\cite{217607} as the carrier audio.

\noindent
\textbf{Software and Hardware.}
We implement ZQ-Attack using the PyTorch framework~\cite{Paszke_PyTorch_An_Imperative_2019}.
The experiments are conducted on a server equipped with 8 NVIDIA 3080Ti GPUs, 2 Intel Xeon Gold 5117 CPUs, and 128 GB RAM, running a 64-bit Ubuntu 18.04 operating system.
In the over-the-air setting, we use a JBL Clip3 speaker to play the audio adversarial examples.
Apple Siri on an iPhone 13 and Amazon Alexa on a second-generation Amazon Echo Dot are used as the target commercial IVC devices.

\noindent
\textbf{Baselines.}
To demonstrate the superior performance of ZQ-Attack, we compare it with several previous works.
In the over-the-line setting, we compare ZQ-Attack with Carlini~\etal~\cite{carlini2018audio},
Occam~\cite{zheng2021black} and KENKU~\cite{291098}.
In the over-the-air setting, we compare ZQ-Attack with NI-Occam~\cite{zheng2021black} and KENKU.
For these baselines, we either utilize their open-source code or re-implement them.
It is noteworthy that the evaluation results of these methods might be inconsistent with the original paper due to periodic updates by manufacturers to their ASR systems.

\noindent
\textbf{Experiment Design.} We evaluate ZQ-Attack on online speech recognition services, commercial IVC devices, and open-source ASRs in Section~\ref{subsec: Evaluation on Online Speech Recognition Services}, Section~\ref{subsec: Evaluation on Commercial IVC Devices}, and Section~\ref{subsec:Evaluation on Open-source ASRs}, respectively.
In Section~\ref{subsec: Impact of Surrogate ASRs}, we explore the impact of surrogate ASRs on ZQ-Attack.
To evaluate the imperceptibility of ZQ-Attack, we conduct a user study in Section~\ref{subsec: user study}.
In Section ~\ref{appendix: Evaluation on a Large Command Set} and Section ~\ref{appendix: Evaluation on Whisper large-v3}, we evaluate ZQ-Attack on a large command set and the latest Whisper large-v3~\cite{whipser-v3}, respectively.

\noindent
\textbf{Ethical Considerations.} We have informed the relevant companies about the potential vulnerability of their ASR systems to our attacks via email.

\begin{table*}[!t]
    \caption{Comparison on online speech recognition services.}
    \label{tab:compare_srs}
    \centering
    \begin{tabular}{c|c|c|c|c|c|c|c}
        \Xhline{1px}
        \multirow{2}{*}{Method}              & \multicolumn{5}{c|}{SRoA $\uparrow$}                                               & \multirow{2}{*}{SNR (dB) $\uparrow$} & \multirow{2}{*}{Query $\downarrow$} \\ \cline{2-6}
                                              & Azure          & Tencent        & Alibaba        & OpenAI         & Average        &                                      &                                     \\ \Xhline{1px}
        Carlini~\etal~\cite{carlini2018audio} & 0/10           & 0/10           & 0/10           & 0/10           & 0/10           & /                                    & 0                                   \\ \hline
        Occam~\cite{zheng2021black}           & 10/10          & 10/10          & 10/10          & 10/10          & 10/10          & 12.54                                & 30000                               \\ \hline
        KENKU~\cite{291098}                   & 10/10          & 8/10           & 0/10           & 9/10           & 6.75/10        & 12.72                                & $>$0                                \\ \hline
        ZQ-Attack                             & \textbf{10/10} & \textbf{10/10} & \textbf{10/10} & \textbf{10/10} & \textbf{10/10} & \textbf{21.91}                       & \textbf{0}                          \\ \Xhline{1px}
    \end{tabular}
\end{table*}

\subsection{Evaluation Metrics}
\label{subsection: evaluation metrics}
We use the success rate of attack (SRoA) as the metric of attack effectiveness.
SRoA is calculated by dividing the number of successfully attacked commands by the total number of commands (\ie 10).
For each target command, if we can generate at least one adversarial example that effectively attacks the target ASR system, we consider the attack on that command as successful.
Note that the adversarial example is considered effective only when its transcription matches exactly the target command.
Any word errors are regarded as a failure, with case sensitivity being disregarded.

To evaluate the imperceptibility of the adversarial examples, we choose signal-to-noise ratio (SNR) as the metric.
SNR is defined as the ratio of the power of a signal (\ie the carrier audio $x$) to the power of a noise (\ie the adversarial perturbation $\delta$) 
in the logarithm scale, and a higher SNR indicates better imperceptibility. The specific calculation method is shown as follows:
\begin{equation}
    \label{equation:snr}
    SNR(\text{dB}) = 10 \cdot \log_{10}\left(\frac{{\Vert x \Vert}_2^2}{{\Vert \delta \Vert}_2^2}\right).
\end{equation}

\subsection{Evaluation on Online Speech Recognition Services}
\label{subsec: Evaluation on Online Speech Recognition Services}
In the over-the-line setting, the results of ZQ-Attack and baseline methods on online speech recognition services are shown in Table~\ref{tab:compare_srs}.
ZQ-Attack successfully generates audio adversarial examples for all target commands on four online speech recognition services, achieving an average SRoA of 100\% and an average SNR of 21.91dB.

For baseline methods, the adversarial examples generated by Carlini~\etal~\cite{carlini2018audio} fail to successfully attack any online speech recognition services, as this method is tailored for the white-box setting.
Compared to Occam, ZQ-Attack achieves comparable effectiveness and better imperceptibility without any queries.
While KENKU still necessitates a small number of queries to search for appropriate hyperparameters tailored to a specific target ASR system, ZQ-Attack attains superior effectiveness and imperceptibility without any queries.
Note that KENKU fails to successfully attack Alibaba in our evaluation.
We speculate that this could be attributed to updates made by Alibaba to its ASR system.

\subsection{Evaluation on Commercial IVC Devices}
\label{subsec: Evaluation on Commercial IVC Devices}

\begin{table}[!t]
    \caption{Comparison on commercial IVC devices.}
    \label{tab:compare_ivc}
    \centering
    \begin{tabular}{c|c|c|c|c}
        \Xhline{1px}
        \multirow{2}{*}{Method}               & \multicolumn{3}{c|}{SRoA $\uparrow$}             & \multirow{2}{*}{SNR (dB) $\uparrow$}   \\
        \cline{2-4}                            & Siri           & Alexa          & Average        &                                        \\ \Xhline{1px}
        NI-Occam~\cite{zheng2021black}         & 4/10           & 5/10           & 4.5/10         & 8.38                                   \\ \hline
        KENKU~\cite{291098}                    & 7/10           & 9/10           & 8/10           & 12.72                                  \\ \hline
        ZQ-Attack                              & \textbf{10/10} & \textbf{10/10} & \textbf{10/10} & \textbf{15.77}                         \\ \Xhline{1px}
    \end{tabular}
\end{table}

\begin{table*}[!t]
    \caption{Evaluation on open-source ASRs.}
    \label{tab:evaluation_os}
    \centering
        \begin{tabular}{c|c|c||c|c|c}
            \Xhline{1px}
            Target ASR        & SRoA  & SNR (dB) & Target ASR                & SRoA  & SNR (dB) \\ \Xhline{1px}
            Jasper            & 10/10 & 13.59    & Conformer-CTC (XL)        & 10/10 & 23.59    \\ \hline
            QuartzNet         & 10/10 & 12.96    & Conformer-Transducer (M)  & 10/10 & 25.34    \\ \hline
            Citrinet (M)      & 10/10 & 14.67    & Conformer-Transducer (L)  & 10/10 & 20.63    \\ \hline
            Citrinet (L)      & 10/10 & 15.89    & Conformer-Transducer (XL) & 10/10 & 21.08    \\ \hline
            ContextNet (M)    & 10/10 & 15.73    & Whisper (base)            & 10/10 & 17.88    \\ \hline
            ContextNet (L)    & 10/10 & 16.87    & Whisper (small)           & 10/10 & 20.76    \\ \hline
            Conformer-CTC (M) & 10/10 & 25.13    & Whisper (medium)          & 10/10 & 23.39    \\ \hline
            Conformer-CTC (L) & 10/10 & 23.51    & Whisper (large)           & 10/10 & 23.74    \\ \Xhline{1px}
        \end{tabular}
\end{table*}

In the over-the-air setting, we generate 10 audio adversarial examples for each target command.  Each adversarial example is played up to three times.
We consider that the attack on a command is successful if at least one adversarial example is effective.

The results of ZQ-Attack and baseline methods on commercial IVC devices are presented in Table~\ref{tab:compare_ivc}.
ZQ-Attack successfully generates audio adversarial examples for all target commands on two commercial IVC devices, achieving an SRoA of 100$\%$ with an average SNR of 15.77dB.
Additionally, an average of 6.6 adversarial examples for each command are effective.

In comparison, the SRoA of ZQ-Attack surpasses KENKU and NI-Occam by 20\% and 55\%.
Concurrently, adversarial examples generated by ZQ-Attack exhibit better imperceptibility, with SNR values exceeding those of KENKU and NI-Occam by 3.05dB and 7.39dB, respectively.

\subsection{Evaluation on Open-source ASRs}
\label{subsec:Evaluation on Open-source ASRs}
To demonstrate the high transferability of audio adversarial examples generated by ZQ-Attack, we additionally evaluate ZQ-Attack on 16 open-source ASR systems.
The evaluation results are shown in Table~\ref{tab:evaluation_os}.
ZQ-Attack successfully generates audio adversarial examples on all 16 open-source ASRs, achieving an average SRoA of 100\% and an average SNR of 19.67dB.
These results show that the audio adversarial examples generated by ZQ-Attack exhibit high transferability, successfully attacking the target open-source ASRs.

Additionally, we observe a potential positive correlation between the recognition performance of open-source ASRs and the success rate of transfer attacks. We speculate that this phenomenon arises due to the better feature extraction capabilities of a more powerful ASR system, facilitating the capture of subtle adversarial perturbations.
The detailed evaluation results and analysis of the correlation can be found in Appendix~\ref{appendix:Evaluation on Open-source ASRs}

\subsection{Impact of Surrogate ASRs}
\label{subsec: Impact of Surrogate ASRs}
ZQ-Attack optimizes the adversarial perturbation on an ordered set of $K$ surrogate ASRs.
To investigate the impact of surrogate ASRs to the performance of ZQ-Attack, we conduct experiments with various values of $K$.
Specifically, we set $K$ to 1, 2, and 4, respectively.
In the cases with only one surrogate ASR, we conduct experiments with four configurations: ContextNet, Citrinet, Conformer-CTC, and Conformer-Transducer (\ie  selecting one from the four available ASRs).
In the cases with two surrogate ASRs, six configurations (\ie combination of two from the four available ASRs) are explored.
Subsequently, we compute the average SRoA and SNR for the three groups of experiments.
In these experiments, we choose one carrier audio and generate audio adversarial examples for 10 target commands.

\begin{figure}[t!]
    \centering
    \begin{subfigure}{0.49\linewidth}
        \centering
        \includegraphics[width=\linewidth]{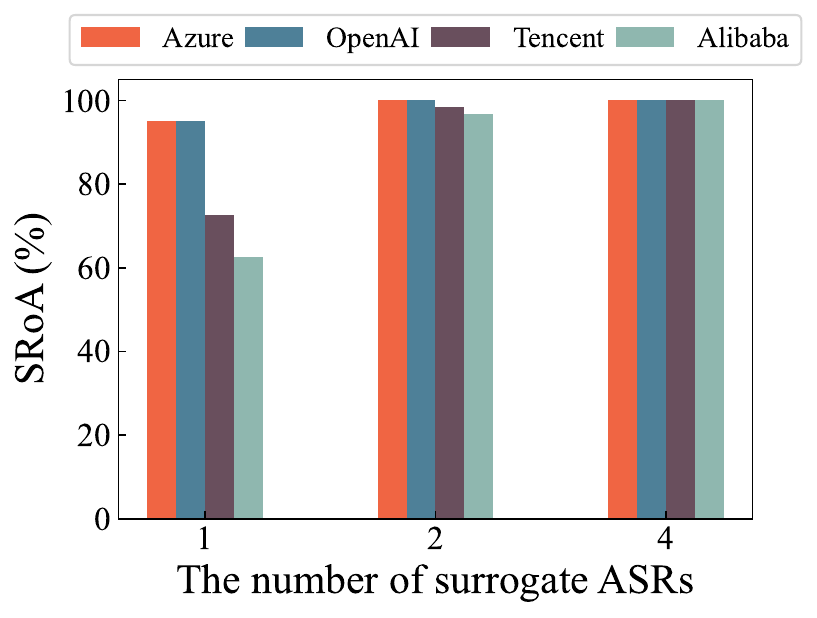}
        \caption{SRoA}
        \label{fig:impact_of_surrogate_ASRs_a}
    \end{subfigure}
    \centering
    \begin{subfigure}{0.49\linewidth}
        \centering
        \includegraphics[width=\linewidth]{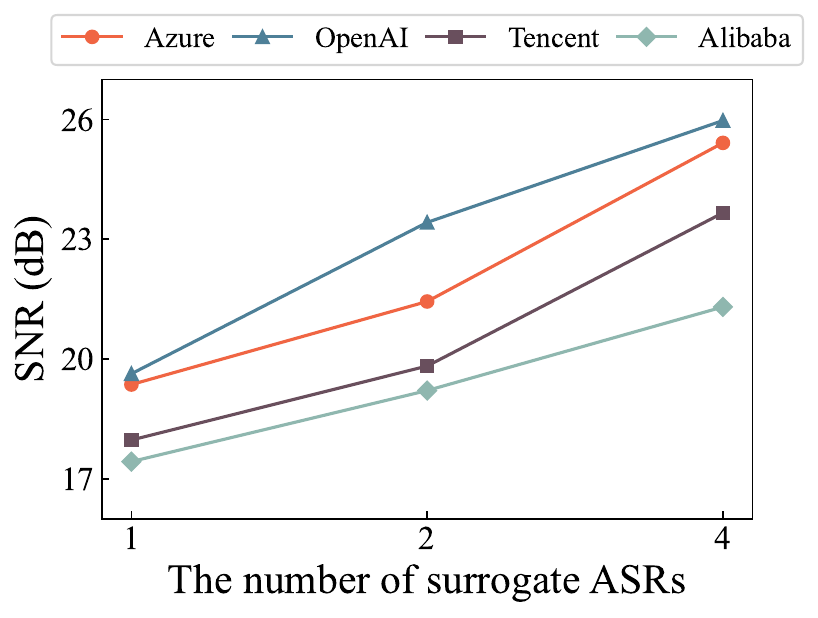}
        \caption{SNR}
        \label{fig:impact_of_surrogate_ASRs_b}
    \end{subfigure}
    \caption{Impact of Surrogate ASRs.}
    \label{fig:impact_of_surrogate_ASRs}

\end{figure}

The results are presented in Figure~\ref{fig:impact_of_surrogate_ASRs}.
It can be observed that with an increase in $K$, both SRoA and SNR exhibit an increasing trend.
For instance, when $K$ is 1, 2, and 4, the average SRoA is 81.25\%, 98.75\%, and 100\%, respectively.
The average SNR for the case involving 4 surrogate ASRs surpasses that of the case with 2 surrogate ASRs by 2.98dB and exceeds the SNR of the case with only 1 surrogate ASR by 5.35dB.
Additionally, it can also be observed that ZQ-Attack successfully performs attacks for most target commands when using 2 surrogate ASRs.
Thus, when computational resources are constrained, and there is a more relaxed requirement for the imperceptibility and effectiveness of the attack, a smaller $K$ can be chosen.
Conversely, a larger $K$ can be used. In the cases with four surrogate ASRs, ZQ-Attack can complete a generation process in approximately 10 minutes using only one NVIDIA 3080Ti GPU.

\captionsetup[subfigure]{skip=0em}

\begin{table*}[!t]
    \caption{User study on human perception.}
    \label{tab:user_srs}
    \centering
    \begin{tabular}{c|c|c|c|c|c|c|c|c}
        \Xhline{1px}
        \multirow{2}{*}{Audio Device} & \multirow{2}{*}{Method}     & \multirow{2}{*}{Normal (\%) $\uparrow$} & \multirow{2}{*}{Noise (\%)} & \multirow{2}{*}{Talking (\%) $\downarrow$} & \multicolumn{2}{c|}{Recognize (\%) $\downarrow$} & \multicolumn{2}{c}{WER (\%) $\uparrow$} \\ \cline{6-9}
                                      &                             &                                         &                             &                                            & \multicolumn{1}{c|}{Once} & Twice                & \multicolumn{1}{c|}{Once} & Twice       \\ \Xhline{1px}
        \multirow{4}{*}{Speaker}      & Songs                       & 92.63                                   & 7.37                        & 0.00                                       & 0.00                      & 0.00                 & 0.00                      & 0.00        \\ \cline{2-9}
                                      & Occam~\cite{zheng2021black} & 5.26                                    & 15.79                       & 78.95                                      & 29.47                     & 35.79                & 57.37                     & 51.24       \\ \cline{2-9}
                                      & KENKU~\cite{291098}         & 0.00                                    & 6.32                        & 93.68                                      & 56.84                     & 63.68                & 30.23                     & 25.96       \\ \cline{2-9}
                                      & ZQ-Attack                   & 13.16                                   & 71.58                       & 15.26                                      & 3.16                      & 3.68                 & 94.95                     & 94.47       \\ \Xhline{1px}

        \multirow{4}{*}{Headphone}    & Songs                       & 92.11                                   & 7.89                        & 0.00                                       & 0.00                      & 0.00                 & 0.00                      & 0.00        \\ \cline{2-9}
                                      & Occam~\cite{zheng2021black} & 2.11                                    & 3.16                        & 94.74                                      & 49.47                     & 54.74                & 38.25                     & 31.32       \\ \cline{2-9}
                                      & KENKU~\cite{291098}         & 0.00                                    & 1.58                        & 98.42                                      & 67.37                     & 70.00                & 19.34                     & 17.77       \\ \cline{2-9}
                                      & ZQ-Attack                   & 5.79                                    & 47.89                       & 46.32                                      & 15.26                     & 22.63                & 77.98                     & 72.41       \\ \Xhline{1px}
    \end{tabular}
\end{table*}

\subsection{User Study}
\label{subsec: user study}
In this section, we recruit human participants to analyze the imperceptibility of the audio adversarial examples and conduct a comparative analysis between ZQ-Attack and the baseline methods (\ie Occam, and KENKU).
This study is carefully designed to mitigate any conceivable risks (psychological, legal, \etc) to the participants, and it is approved by the institutional review board (IRB).
The target commands are common household phrases (\eg``call my wife'') to minimize discomfort and the audio volume is normalized to maintain it below a safe threshold, preventing any risk of hearing damage.
Besides, our study does not collect any private information from the participants, and all data will be deleted upon the completion of the study.

Our test audios comprise both normal audios and audio adversarial examples.
The normal audios consist of 10 songs, and the audio adversarial examples include 10 instances from each method.
In the user study, each participant listens to all test audios.
Following the first listening, participants are provided with three choices: ``normal audio'', ``noisy audio'', and ``audio with background speech''.
In cases where participants select the ``audio with background speech'' option, they are required to provide the content of the perceived speech, followed by a second round of listening and transcription of the same test audio.
Specifically, participants first listen to the test audios through speakers (\eg MacBook Pro Speaker).
Then, to eliminate environmental interference, participants listen to the audios again through noise-canceling headphones (\eg Sony WH-1000XM5).

We gather data from a total of 38 participants, consisting of 20 males and 18 females.
Among this group, 10 participants are below the age of 22, 18 participants are aged 22 to 24, and 10 participants are 25 or older.
All participants are proficient in both spoken and written English, hold at least a bachelor's degree, and have normal hearing.
The results are presented in Table~\ref{tab:user_srs}.
When using the speaker as the audio device, 13.16\% of participants select our adversarial examples as ``normal audio'', while 71.58\% of participants select them as ``noisy audio''.
Although 15.26\% of participants select our adversarial examples as ``audio with background speech,'' only 3.68\% of commands are recognized even after a second round of listening and transcription.
In a more challenging configuration that uses headphones to play the audios, only 22.63\% of the commands are recognized even after a second round of listening and transcription.
Additionally, for the audios played using speakers and headphones, the average word error rate (WER) between user recognition results and the actual target commands is 94.47\% and 72.41\%, respectively.
Compared with other methods, ZQ-Attack attains superior imperceptibility.

Furthermore, we observed that the imperceptibility of different target commands varies. The command \textit{open the door} is the most easily perceived, while \textit{send a text} is the hardest to perceive. This disparity may be attributed to the varying phonemes in different command audios, with vowel phonemes containing more energy and thus being more easily audible~\cite{10.1145/3548606.3560660,yu2023smack}.

\begin{table*}[!t]
    \caption{Evaluation on a large command set.}
    \label{tab:evaluation_large}
    \centering
        \begin{tabular}{c|c|c|c|c|c|c|c|c}
            \Xhline{1px}
            \multirow{2}{*}{Command}    & \multicolumn{2}{c|}{Azure}             & \multicolumn{2}{c|}{Tencent}           & \multicolumn{2}{c|}{Alibaba}           & \multicolumn{2}{c}{OpenAI}             \\ \cline{2-9}
                                        & \multicolumn{1}{c|}{Attack} & SNR (dB) & \multicolumn{1}{c|}{Attack} & SNR (dB) & \multicolumn{1}{c|}{Attack} & SNR (dB) & \multicolumn{1}{c|}{Attack} & SNR (dB) \\ \Xhline{1px}

            ask me a question           & \ding{51}                   & 23.51    & \ding{51}                   & 28.66    & \ding{51}                   & 26.31    & \ding{51}                   & 28.73    \\ \hline
            clear notification          & \ding{51}                   & 26.52    & \ding{51}                   & 21.28    & \ding{51}                   & 20.05    & \ding{51}                   & 26.52    \\ \hline
            close the shades            & \ding{51}                   & 26.54    & \ding{51}                   & 26.24    & \ding{51}                   & 26.54    & \ding{51}                   & 26.86    \\ \hline
            find a hotel                & \ding{51}                   & 25.78    & \ding{51}                   & 24.63    & \ding{51}                   & 20.26    & \ding{51}                   & 28.21    \\ \hline
            good morning                & \ding{51}                   & 19.65    & \ding{51}                   & 18.96    & \ding{51}                   & 27.90    & \ding{51}                   & 25.76    \\ \hline
            I have a secret to tell you & \ding{51}                   & 27.21    & \ding{51}                   & 27.21    & \ding{51}                   & 27.21    & \ding{51}                   & 27.21    \\ \hline
            I need help                 & \ding{51}                   & 27.23    & \ding{51}                   & 26.64    & \ding{51}                   & 20.17    & \ding{51}                   & 28.54    \\ \hline
            open the box                & \ding{51}                   & 18.48    & \ding{51}                   & 26.34    & \ding{51}                   & 18.48    & \ding{51}                   & 27.54    \\ \hline
            read a book                 & \ding{51}                   & 20.77    & \ding{51}                   & 25.81    & \ding{51}                   & 17.27    & \ding{51}                   & 27.71    \\ \hline
            record a video              & \ding{51}                   & 21.38    & \ding{51}                   & 20.51    & \ding{51}                   & 21.38    & \ding{51}                   & 21.38    \\ \hline
            reset password              & \ding{51}                   & 22.70    & \ding{51}                   & 21.69    & \ding{51}                   & 19.91    & \ding{51}                   & 22.70    \\ \hline
            show me my message          & \ding{51}                   & 27.12    & \ding{51}                   & 27.54    & \ding{51}                   & 23.98    & \ding{51}                   & 27.54    \\ \hline
            show me the money           & \ding{51}                   & 27.94    & \ding{51}                   & 27.96    & \ding{51}                   & 25.75    & \ding{51}                   & 27.96    \\ \hline
            start recording             & \ding{51}                   & 27.03    & \ding{51}                   & 20.42    & \ding{51}                   & 19.87    & \ding{51}                   & 27.19    \\ \hline
            tell me a story             & \ding{51}                   & 27.89    & \ding{51}                   & 27.89    & \ding{51}                   & 24.34    & \ding{51}                   & 27.94    \\ \hline
            turn off the fan            & \ding{51}                   & 19.98    & \ding{51}                   & 15.25    & \ding{51}                   & 19.98    & \ding{51}                   & 19.98    \\ \hline
            turn on the TV              & \ding{51}                   & 25.55    & \ding{51}                   & 17.71    & \ding{51}                   & 17.18    & \ding{51}                   & 25.55    \\ \hline
            watch TV                    & \ding{51}                   & 26.32    & \ding{51}                   & 26.32    & \ding{51}                   & 19.20    & \ding{51}                   & 26.32    \\ \hline
            what time is it             & \ding{51}                   & 26.22    & \ding{51}                   & 25.39    & \ding{51}                   & 25.39    & \ding{51}                   & 28.20    \\ \hline
            where is my car             & \ding{51}                   & 27.24    & \ding{51}                   & 23.28    & \ding{51}                   & 25.76    & \ding{51}                   & 27.24    \\ \Xhline{1px}
            Average                     & 20/20                       & 24.75    & 20/20                       & 23.99    & 20/20                       & 22.35    & 20/20                       & 26.45    \\ \Xhline{1px}
        \end{tabular}
\end{table*}

\subsection{Evaluation on a Larger Command Set}
\label{appendix: Evaluation on a Large Command Set}

To thoroughly evaluate the effectiveness of ZQ-Attack, we conduct experiments using a larger command set.
This set includes a total of 20 commands: \textit{ask me a question}, \textit{clear notification}, \textit{close the shades} \textit{find a hotel}, \textit{good morning}, \textit{I have a secret to tell you}, \textit{I need help}, \textit{open the box}, \textit{read a book}, \textit{record a video}, \textit{reset password}, \textit{show me my message}, \textit{show me the money}, \textit{start recording}, \textit{tell me a story}, \textit{turn off the fan}, \textit{turn on the TV}, \textit{watch TV}, \textit{what time is it} \textit{where is my car}.

We choose a carrier audio and generate audio adversarial examples for each of these target commands.
As illustrated in Table~\ref{tab:evaluation_large}, ZQ-Attack can generate effective audio adversarial examples for each target command in this set, achieving an average SRoA of 100\%.
On Azure, Tencent, Alibaba, and OpenAI, ZQ-Attack attains average SNR values of 24.75dB, 23.99dB, 22.35dB, and 26.45dB, respectively.

\subsection{Evaluation on Whisper large-v3}
\label{appendix: Evaluation on Whisper large-v3}
Whisper \cite{whisper} is a popular ASR system trained on a diverse audio dataset comprising 680,000 hours of audio, achieving state-of-the-art multilingual recognition performance.
Evaluating ZQ-Attack on Whisper can more comprehensively demonstrate its effectiveness on state-of-the-art ASR systems.
Since the models of Whisper come in diverse sizes, we also conduct experiments on these various models.
In Section~\ref{subsec:Evaluation on Open-source ASRs}, we evaluate ZQ-Attack utilizing the base, small, medium, and large models as open-source ASRs.
As the OpenAI API employs the large-v2 model, we have evaluated the performance of ZQ-Attack on the large-v2 model in Section~\ref{subsec: Evaluation on Online Speech Recognition Services}. 

Recently, OpenAI introduced the latest large-v3 model \cite{whipser-v3}, which surpasses the performance of the large-v2 model.
We have also evaluated ZQ-Attack on the latest model.
As indicated in Table~\ref{tab:evaluation_whisper3}, ZQ-Attack can successfully generate effective adversarial examples for all 10 target commands, achieving an average SNR of 23.49dB.
These results demonstrate the capability of ZQ-Attack to generate effective and imperceptible audio adversarial examples on the most advanced ASR systems.

\begin{table}[!t]
    \caption{Evaluation on the latest Whisper large-v3.}
    \label{tab:evaluation_whisper3}
    \centering
    \begin{tabular}{c|c|c}
        \Xhline{1px}
        \multirow{2}{*}{Command} & \multicolumn{2}{c}{Whisper large-v3} \\ \cline{2-3}
                                 & Attack   & SNR (dB)       \\ \Xhline{1px}
        call my wife             & \ding{51}  & 21.28          \\ \hline
        make it warmer           & \ding{51}  & 17.31          \\ \hline
        navigate to my home      & \ding{51}  & 24.35          \\ \hline
        open the door            & \ding{51} & 24.46          \\ \hline
        open the website         & \ding{51}  & 26.75          \\ \hline
        play music               & \ding{51} & 24.76          \\ \hline
        send a text              & \ding{51}   & 24.54          \\ \hline
        take a picture           & \ding{51}  & 27.23          \\ \hline
        turn off the light       & \ding{51}  & 26.29          \\ \hline
        turn on airplane mode    & \ding{51}  & 17.95          \\ \Xhline{1px}
        Average                  & 10/10      & 23.49          \\ \Xhline{1px}
    \end{tabular}
\end{table}

\section{Related Work}

\begin{table*}[!t]
    \caption{SRoA of ZQ-Attack on online speech recognition services with defenses.}
    \label{tab:defense}
    \centering
    \begin{tabular}{c|c|c|c|c|c|c}
        \Xhline{1px}
        Defenses                             & Setting         & Azure & Tencent & Alibaba & OpenAI & Average \\ \Xhline{1px}
        \multirow{3}{*}{Local Smoothing}     & $h=1$           & 10/10 & 10/10   & 10/10   & 10/10  & 10/10   \\ \cline{2-7}
                                             & $h=2$           & 10/10 & 10/10   & 10/10   & 10/10  & 10/10   \\ \cline{2-7}
                                             & $h=3$           & 10/10 & 10/10   & 9/10    & 10/10  & 9.75/10 \\ \Xhline{1px}
        \multirow{3}{*}{Downsampling}        & $f_{low}=$~14kHz & 10/10 & 10/10   & 10/10   & 10/10  & 10/10   \\ \cline{2-7}
                                             & $f_{low}=$~12kHz & 10/10 & 10/10   & 9/10    & 9/10   & 9.5/10  \\ \cline{2-7}
                                             & $f_{low}=$~10kHz & 10/10 & 10/10   & 9/10    & 9/10   & 9.5/10  \\ \Xhline{1px}
        \multirow{3}{*}{Temporal Dependency} & $k=0.2$         & 10/10 & 10/10   & 10/10   & 10/10  & 10/10   \\ \cline{2-7}
                                             & $k=0.5$         & 10/10 & 10/10   & 10/10   & 10/10  & 10/10   \\ \cline{2-7}
                                             & $k=0.8$         & 10/10 & 10/10   & 10/10   & 10/10  & 10/10   \\ \Xhline{1px}
        \multirow{3}{*}{MVP-EARS}            & $m=2$           & 10/10 & 10/10   & 10/10   & 10/10  & 10/10   \\ \cline{2-7}
                                             & $m=3$           & 10/10 & 10/10   & 10/10   & 10/10  & 10/10   \\ \cline{2-7}
                                             & $m=4$           & 10/10 & 10/10   & 10/10   & 10/10  & 10/10   \\ \Xhline{1px}
    \end{tabular}
\end{table*}

\noindent
\textbf{Audio Adversarial Attacks on White-box ASR Systems.}
In recent years, extensive studies have focused on audio adversarial attacks on ASR systems.
Carlini \etal~\cite{carlini2018audio} were the first to generate targeted audio adversarial examples on the white-box DeepSpeech model.
Concurrently, Commandersong~\cite{217607} demonstrated successful attacks on the Kaldi model by embedding malicious commands into songs.
This approach also enabled physical attacks but had stringent limitations, such as requiring specific speakers and recording devices.
Qin~\etal~\cite{qin2019imperceptible} and Schönherr~\etal~\cite{schonherr2018adversarial} contributed to improving the imperceptibility of generated adversarial examples by incorporating the psychoacoustic model into the audio adversarial example generation process.

\noindent
\textbf{Audio Adversarial Attacks on Black-box ASR Systems.}
Despite the success of the aforementioned methods in generating audio adversarial examples on white-box ASR systems, their reliance on the gradient information of the target model limits their applicability for attacking black-box ASR systems.
To address this challenge, Taori~\etal~\cite{taori2019targeted} proposed employing gradient estimation and genetic algorithms to achieve black-box attacks, but their method had a relatively low attack success rate.
Following this work, SGEA~\cite{wang2020towards} improved the attack success rate and reduced the number of queries by employing selective gradient estimation techniques.
Nevertheless, generating a single audio adversarial example still required approximately 100,000 queries.
Devil's Whisper~\cite{chen2020devil} significantly improved the attack success rate by using a surrogate model. However, it relied on confidence scores returned by the target ASR system, which often be unavailable in real-world scenarios.
Occam~\cite{zheng2021black} employed cooperative co-evolution and the CMA evolution strategy~\cite{hansen2016cma}, eliminating the need for confidence scores.
Recently, KENKU~\cite{291098} optimized the acoustic feature loss based on MFCC and imperceptibility loss simultaneously to generate relatively stealthy audio adversarial examples on black-box ASR systems. Although these methods have achieved improved performance, their dependence on querying the target system continues to limit their practicality.

To generate audio adversarial examples without the need for queries, researchers have proposed transfer-based attacks that can generate adversarial examples capable of attacking different models.
NI-Occam~\cite{zheng2021black} utilized fine-tuned Kaldi models to exclusively launch attacks on IVC devices that are more sensitive to commands, but its attack success rate remains relatively low.
TransAudio~\cite{qi2023transaudio} presented a two-stage framework and a score-matching-based optimization strategy to achieve word-level adversarial attack, but its target transcription is constrained by the carrier audio. In the field of image adversarial attacks, previous work~\cite{brown2017adversarial,liu2016delving} proposed utilizing the ensemble method to improve the transferability of adversarial examples. Inspired by this, we design a sequential ensemble optimization algorithm to generate adversarial examples using diverse surrogate ASRs.

\section{Discussion}
\subsection{Defenses against ZQ-Attack}
We evaluate ZQ-Attack with several state-of-the-art defense methods against audio adversarial attacks, including local smoothing, downsampling, temporal dependency~\cite{yang2018characterizing}, and MVP-EARS~\cite{zeng2019multiversion}. The results are presented in Table~\ref{tab:defense}.

\noindent
\textbf{Local Smoothing.}
Local smoothing renders audio adversarial attacks ineffective by applying a sliding window with a median filter to the adversarial examples.
Given the length of the sliding window, denoted as $h$, the value of an audio sample point is replaced by the average values of itself and the $h$ sample points before and after it.
To evaluate the robustness of ZQ-Attack against local smoothing, we set $h$ to 1, 2, and 3, respectively.
The results in Table~\ref{tab:defense} show that local smoothing has minimal impact on ZQ-Attack.
Across different settings, ZQ-Attack consistently achieves an SRoA higher than 97.5\%.

\noindent
\textbf{Downsampling.}
This method involves downsampling the original audio to a lower sampling rate $f_{low}$ and subsequently upsampling it to the original sampling rate, which causes a loss of high-frequency information from the original audio.
Therefore, if the high-frequency information in the adversarial perturbation is lost, the attack might fail.
To assess the robustness of ZQ-Attack against downsampling, we set $f_{low}$ to 14kHz, 12kHz, and 10kHz.
The results in Table~\ref{tab:defense} show that
ZQ-Attack has great robustness to downsampling, achieving an SRoA higher than 95\% under different settings.

\noindent
\textbf{Temporal Dependency.}
The inherent temporal dependency in audio data can be leveraged to detect audio adversarial examples \cite{yang2018characterizing}.
Specifically, audio adversarial examples can be identified by comparing the transcription of the first $k$ part of the audio with the first $k$ part of the transcription of the entire audio,  where $k$ is a ratio between 0 and 1.
If the consistency between them is low, the audio can be considered an adversarial example; otherwise, it is accepted as normal.
To evaluate the robustness of ZQ-Attack against the temporal dependency-based defense, we set $k$ to 0.2, 0.5, and 0.8.
The results in Table~\ref{tab:defense} demonstrate that the audio adversarial examples generated by ZQ-Attack exhibit strong resilience to temporal dependency, as all attacks are successful under different settings.

\noindent
\textbf{MVP-EARS \cite{zeng2019multiversion}.}
MVP-EARS utilizes multiple ASR systems to detect audio adversarial examples. Due to the limited transferability of most prior audio adversarial example generation methods, different ASR systems may produce significantly different transcriptions for the same audio adversarial example. Therefore, multiple ASR systems can be used to transcribe the same audio. If their transcripts differ, the audio can be considered an adversarial example.

To evaluate the robustness of ZQ-Attack against MVP-EARS, we set $m$ to 2, 3, and 4, where $m$ is the number of ASR systems used.
The results in Table~\ref{tab:defense} show that the audio adversarial examples generated by ZQ-Attack exhibit good robustness to MVP-EARS, with all target commands successfully attacked under different settings.
This is attributed to the fact that ZQ-Attack does not generate audio adversarial examples customized for a specific ASR but crafts transferable adversarial examples on diverse surrogate ASRs.

\subsection{Limitations and Future Work}

While ZQ-Attack leverages diverse surrogate ASRs to achieve transferable audio adversarial attacks in the zero-query black-box setting, it incurs a higher computation cost (\eg GPU memory).
However, considering that the audio adversarial examples generated by ZQ-Attack are effective on multiple ASR systems and save the cost of queries, we consider that the additional cost is acceptable.
Another limitation is that although the audio adversarial examples generated by ZQ-Attack exhibit better imperceptibility compared to prior work in the over-the-air setting, they may still be detected by humans.
We leave enhancing the imperceptibility of audio adversarial examples in the over-the-air setting for future work.

\section{Conclusion}
We proposed ZQ-Attack, a transfer-based adversarial attack on ASR systems in the zero-query black-box scenario.
By summarizing and categorizing the modern ASR systems, we first selected a diverse set of surrogate ASRs for generating adversarial examples.
Then, we employed an adaptive search algorithm to initialize the adversarial perturbations with a scaled target command audio, ensuring its effectiveness and imperceptibility.
Subsequently, we designed a novel sequential ensemble optimization algorithm to optimize the adversarial perturbations using the selected surrogate ASRs.
Our experimental results indicate that ZQ-attack achieves successful attacks on 4 online speech recognition services and 16 open-source ASRs in the over-the-line setting and attacks 2 commercial IVC devices in the over-the-air setting. This demonstrates a significant improvement in the practicality of audio adversarial attacks compared to prior methods.

\begin{acks}
We thank the anonymous reviewers for their helpful and valuable feedback. This work was partially supported by the National Key R\&D Program of China 2023YFE0209800, NSFC under Grants U20B2049, U21B2018, 62302344, 62132011, 62161160337, and Shaanxi Province Key Industry Innovation Program 2021ZDLGY01-02.
\end{acks}

\bibliographystyle{ACM-Reference-Format}
\balance
\bibliography{ref}

\appendix
\begin{table*}[!t]
    \caption{Detailed results of evaluation on online speech recognition services.}
    \label{tab:evaluation_srs}
    \centering
        \begin{tabular}{c|c|c|c|c|c|c|c|c}
            \Xhline{1.5px}
            \multirow{2}{*}{Command} & \multicolumn{2}{c|}{Azure} & \multicolumn{2}{c|}{Tencent}      & \multicolumn{2}{c|}{Alibaba}     & \multicolumn{2}{c}{OpenAI}                                                                   \\ \cline{2-9}
                                     & Attack    & SNR (dB)       & Attack                & SNR (dB)  & Attack    & SNR (dB)             & Attack    & SNR (dB)                                                                         \\ \Xhline{1px}
            call my wife             & \ding{51} & 18.41          & \ding{51}             & 19.88     & \ding{51} & 16.89                & \ding{51} & 17.44                                                                            \\ \hline
            make it warmer           & \ding{51} & 20.35          & \ding{51}             & 18.18     & \ding{51} & 16.50                & \ding{51} & 15.41                                                                            \\ \hline
            navigate to my home      & \ding{51} & 20.85          & \ding{51}             & 23.42     & \ding{51} & 20.65                & \ding{51} & 20.35                                                                            \\ \hline
            open the door            & \ding{51} & 21.63          & \ding{51}             & 23.09     & \ding{51} & 20.73                & \ding{51} & 17.67                                                                            \\ \hline
            open the website         & \ding{51} & 21.86          & \ding{51}             & 26.67     & \ding{51} & 23.94                & \ding{51} & 23.55                                                                            \\ \hline
            play music               & \ding{51} & 25.47          & \ding{51}             & 26.51     & \ding{51} & 20.26                & \ding{51} & 20.97                                                                            \\ \hline
            send a text              & \ding{51} & 25.60          & \ding{51}             & 23.28     & \ding{51} & 25.79                & \ding{51} & 23.18                                                                            \\ \hline
            take a picture           & \ding{51} & 26.85          & \ding{51}             & 27.77     & \ding{51} & 25.00                & \ding{51} & 26.48                                                                            \\ \hline
            turn off the light       & \ding{51} & 25.08          & \ding{51}             & 27.94     & \ding{51} & 26.16                & \ding{51} & 24.28                                                                            \\ \hline
            turn on airplane mode    & \ding{51} & 17.44          & \ding{51}             & 17.79     & \ding{51} & 16.53                & \ding{51} & 16.75                                                                            \\ \Xhline{1px}
            Average                  & \ding{51} & 22.35          & \ding{51}             & 23.45     & \ding{51} & 21.24                & \ding{51} & 20.61                                                                            \\ \Xhline{1px}
        \end{tabular}
        % }
\end{table*}

\section{Evaluation on Online Speech Recognition Services}
\label{appendix: Evaluation on Online Speech Recognition Services}

The detailed evaluation results of ZQ-Attack on online speech recognition services are presented in Table~\ref{tab:evaluation_srs}. For the 10 target commands, ZQ-Attack can generate effective adversarial examples on 4 online speech recognition services.
The results demonstrate that ZQ-Attack can successfully generate effective and imperceptible audio adversarial examples on these online speech recognition services.

\begin{table}[t!]
\caption{Detailed results of evaluation on commercial IVC devices.}
\label{tab:evaluation_ivc}
\resizebox{\linewidth}{!}{
\begin{tabular}{c|c|c|c|c|c|c}
\Xhline{1.5px}
\multirow{2}{*}{Command} & \multicolumn{2}{c|}{NI-Occam}     & \multicolumn{2}{c|}{KENKU}        & \multicolumn{2}{c}{ZQ-Attack}     \\ \cline{2-7} 
                         & Siri & Alexa & Siri & Alexa & Siri & Alexa \\ \Xhline{1.5px}
call my wife             & \ding{51}    &   \ding{51}    &  \ding{51}   &    \ding{51}   & \ding{51}    &  \ding{51}     \\ \hline
make it warmer           & \ding{55}    &   \ding{55}    &  \ding{55}   &    \ding{51}   &    \ding{51}    &  \ding{51}       \\ \hline
navigate to my home      &  \ding{51}   &   \ding{51}    &  \ding{51}   &   \ding{51}    &     \ding{51}    &  \ding{51}       \\ \hline
open the door            &  \ding{55}   &   \ding{55}    &  \ding{51}   &    \ding{51}   &    \ding{51}    &  \ding{51}       \\ \hline
open the website         &  \ding{55}   &  \ding{51}     &  \ding{51}   &   \ding{51}    &     \ding{51}    &  \ding{51}       \\ \hline
play music               &  \ding{51}   &   \ding{51}    &   \ding{51}  &   \ding{51}    &     \ding{51}    &  \ding{51}       \\ \hline
send a text              &  \ding{55}   &   \ding{55}    &  \ding{55}   &   \ding{55}    &     \ding{51}    &  \ding{51}       \\ \hline
take a picture           &   \ding{55}  &   \ding{55}    &  \ding{55}   &   \ding{51}    &     \ding{51}    &  \ding{51}       \\ \hline
turn off the light       &  \ding{51}   &    \ding{51}   &  \ding{51}  &    \ding{51}   &     \ding{51}    &  \ding{51}       \\ \hline
turn on airplane mode    &  \ding{55}   &  \ding{55}     &  \ding{51}   &   \ding{51}    &     \ding{51}    &  \ding{51}       \\ \hline
Average                  & 4/10    & 5/10      & 7/10    &  9/10     &  10/10   & 10/10       \\ \Xhline{1.5px}
\end{tabular}
}
\end{table}

\section{Evaluation on Commercial IVC Devices}
\label{appendix: Evaluation on Commercial IVC Devices}
The detailed evaluation results on commercial IVC devices are presented in Table~\ref{tab:evaluation_ivc}. The results demonstrate that ZQ-Attack successfully attacks all 10 target commands, outperforming both KENKU and NI-Occam.
Moreover, we observe that the effectiveness of these methods can vary significantly for different commands. For example, for the command \textit{play music}, all methods successfully execute the attack. However, for the command \textit{make it warmer}, both KENKU and NI-Occam fail to achieve success.

\section{Evaluation on Open-source ASRs}
\label{appendix:Evaluation on Open-source ASRs}
We introduce the transfer rate (TR) to evaluate the transferability of audio adversarial examples generated by ZQ-Attack to open-source ASR systems.
For a specified target command and target open-source ASR system, the TR represents the success rate of the transfer attack, \ie the ratio of adversarial examples within $X^\prime$ that successfully attack the target open-source ASR system.
A higher TR indicates better transferability of the audio adversarial examples generated by ZQ-Attack to the target open-source ASR system.
The results are presented in Table~\ref{tab:detail_os}. ZQ-Attacks obtains an average TR of 65.19\% and an SNR of 19.67dB on 16 open-source ASRs.
These results indicate that ZQ-Attack can generate audio adversarial examples that effectively attack a diverse range of open-source ASRs.

\begin{table*}[!t]
    \caption{Evaluation on open-source ASRs.}
    \label{tab:detail_os}
    \centering
    \resizebox{0.73\linewidth}{!}{
        \begin{tabular}{c|c|c|c|c|c|c|c|c|c|c|c|c}
            \Xhline{1.5px}
            \multirow{2}{*}{Command} & \multicolumn{3}{c|}{Jasper}             & \multicolumn{3}{c|}{QuartzNet}                & \multicolumn{3}{c|}{Citrinet (M)}             & \multicolumn{3}{c}{Citrinet (L)}              \\ \cline{2-13}
                                     & Attack    & TR (\%) & SNR (dB)          & Attack    & TR (\%) & SNR (dB)                & Attack    & TR (\%) & SNR (dB)                & Attack    & TR (\%) & SNR (dB)                \\ \Xhline{1px}
            call my wife             & \ding{51} & 31.73   & 9.97              & \ding{51} & 33.35   & 9.90                    & \ding{51} & 24.08   & 11.60                   & \ding{51} & 50.28   & 13.46                   \\ \hline
            make it warmer           & \ding{51} & 18.21   & 9.43              & \ding{51} & 27.84   & 9.94                    & \ding{51} & 12.50   & 9.64                    & \ding{51} & 38.20   & 13.00                   \\ \hline
            navigate to my home      & \ding{51} & 27.35   & 12.28             & \ding{51} & 15.35   & 11.52                   & \ding{51} & 20.73   & 13.88                   & \ding{51} & 38.42   & 14.71                   \\ \hline
            open the door            & \ding{51} & 28.60   & 13.19             & \ding{51} & 23.30   & 12.96                   & \ding{51} & 11.15   & 13.33                   & \ding{51} & 27.19   & 14.54                   \\ \hline
            open the website         & \ding{51} & 19.06   & 15.13             & \ding{51} & 12.63   & 14.72                   & \ding{51} & 13.08   & 17.98                   & \ding{51} & 23.32   & 15.59                   \\ \hline
            play music               & \ding{51} & 38.72   & 14.04             & \ding{51} & 32.41   & 13.41                   & \ding{51} & 29.91   & 16.20                   & \ding{51} & 55.16   & 17.54                   \\ \hline
            send a text              & \ding{51} & 44.28   & 18.50             & \ding{51} & 22.50   & 15.94                   & \ding{51} & 2.58    & 15.69                   & \ding{51} & 61.94   & 19.70                   \\ \hline
            take a picture           & \ding{51} & 20.42   & 19.98             & \ding{51} & 16.25   & 17.99                   & \ding{51} & 21.92   & 22.79                   & \ding{51} & 36.07   & 22.07                   \\ \hline
            turn off the light       & \ding{51} & 14.78   & 17.55             & \ding{51} & 11.85   & 16.36                   & \ding{51} & 13.03   & 18.73                   & \ding{51} & 26.31   & 20.66                   \\ \hline
            turn on airplane mode    & \ding{51} & 7.27    & 5.82              & \ding{51} & 24.33   & 6.86                    & \ding{51} & 10.59   & 6.87                    & \ding{51} & 21.75   & 7.59                    \\ \Xhline{1px}
            Average                  & 10/10     & 25.04   & 13.59             & 10/10     & 21.98   & 12.96                   & 10/10     & 15.96   & 14.67                   & 10/10     & 37.86   & 15.89                   \\ \Xhline{1.5px}

            \multirow{2}{*}{Command} & \multicolumn{3}{c|}{ContextNet (M)}     & \multicolumn{3}{c|}{ContextNet (L)}           & \multicolumn{3}{c|}{Conformer-CTC (M)}        & \multicolumn{3}{c}{Conformer-CTC (L)}         \\ \cline{2-13}
                                     & Attack    & TR (\%) & SNR (dB)          & Attack    & TR (\%) & SNR (dB)                & Attack    & TR (\%) & SNR (dB)                & Attack    & TR (\%) & SNR (dB)                \\ \Xhline{1px}
            call my wife             & \ding{51} & 75.56   & 15.37             & \ding{51} & 69.31   & 14.18                   & \ding{51} & 100.00  & 23.06                   & \ding{51} & 93.02   & 20.66                   \\ \hline
            make it warmer           & \ding{51} & 62.58   & 13.03             & \ding{51} & 65.07   & 13.01                   & \ding{51} & 91.76   & 23.17                   & \ding{51} & 85.89   & 17.71                   \\ \hline
            navigate to my home      & \ding{51} & 53.39   & 14.95             & \ding{51} & 56.53   & 16.71                   & \ding{51} & 96.57   & 25.52                   & \ding{51} & 80.47   & 19.88                   \\ \hline
            open the door            & \ding{51} & 39.96   & 14.02             & \ding{51} & 50.64   & 15.40                   & \ding{51} & 99.71   & 26.84                   & \ding{51} & 94.56   & 25.33                   \\ \hline
            open the website         & \ding{51} & 31.78   & 15.56             & \ding{51} & 49.42   & 17.73                   & \ding{51} & 99.34   & 26.75                   & \ding{51} & 87.89   & 25.05                   \\ \hline
            play music               & \ding{51} & 46.42   & 14.78             & \ding{51} & 61.80   & 17.81                   & \ding{51} & 99.05   & 26.84                   & \ding{51} & 99.72   & 26.85                   \\ \hline
            send a text              & \ding{51} & 49.05   & 19.44             & \ding{51} & 56.27   & 18.73                   & \ding{51} & 93.02   & 25.37                   & \ding{51} & 98.58   & 25.91                   \\ \hline
            take a picture           & \ding{51} & 27.93   & 19.76             & \ding{51} & 37.87   & 21.81                   & \ding{51} & 99.91   & 27.78                   & \ding{51} & 97.82   & 27.78                   \\ \hline
            turn off the light       & \ding{51} & 42.07   & 18.46             & \ding{51} & 55.41   & 20.58                   & \ding{51} & 98.01   & 27.89                   & \ding{51} & 96.97   & 27.84                   \\ \hline
            turn on airplane mode    & \ding{51} & 67.30   & 11.96             & \ding{51} & 52.77   & 12.73                   & \ding{51} & 98.66   & 18.06                   & \ding{51} & 83.09   & 18.06                   \\ \Xhline{1px}
            Average                  & 10/10     & 49.60   & 15.73             & 10/10     & 55.51   & 16.87                   & 10/10     & 97.60   & 25.13                   & 10/10     & 91.80   & 23.51                   \\ \Xhline{1.5px}

            \multirow{2}{*}{Command} & \multicolumn{3}{c|}{Conformer-CTC (XL)} & \multicolumn{3}{c|}{Conformer-Transducer (M)} & \multicolumn{3}{c|}{Conformer-Transducer (L)} & \multicolumn{3}{c}{Conformer-Transducer (XL)} \\ \cline{2-13}
                                     & Attack    & TR (\%) & SNR (dB)          & Attack    & TR (\%) & SNR (dB)                & Attack    & TR (\%) & SNR (dB)                & Attack    & TR (\%) & SNR (dB)                \\ \Xhline{1px}
            call my wife             & \ding{51} & 96.61   & 22.02             & \ding{51} & 97.99   & 22.99                   & \ding{51} & 96.19   & 20.58                   & \ding{51} & 83.21   & 20.10                   \\ \hline
            make it warmer           & \ding{51} & 84.56   & 18.87             & \ding{51} & 97.82   & 24.77                   & \ding{51} & 85.14   & 18.37                   & \ding{51} & 43.34   & 18.48                   \\ \hline
            navigate to my home      & \ding{51} & 97.54   & 25.47             & \ding{51} & 99.76   & 25.52                   & \ding{51} & 84.51   & 22.05                   & \ding{51} & 93.03   & 23.30                   \\ \hline
            open the door            & \ding{51} & 91.13   & 25.90             & \ding{51} & 100.00  & 26.84                   & \ding{51} & 54.29   & 16.05                   & \ding{51} & 60.62   & 18.73                   \\ \hline
            open the website         & \ding{51} & 87.18   & 24.17             & \ding{51} & 100.00  & 26.76                   & \ding{51} & 41.65   & 23.85                   & \ding{51} & 36.62   & 21.09                   \\ \hline
            play music               & \ding{51} & 97.26   & 26.85             & \ding{51} & 98.77   & 26.85                   & \ding{51} & 90.61   & 26.30                   & \ding{51} & 59.51   & 22.27                   \\ \hline
            send a text              & \ding{51} & 92.70   & 24.39             & \ding{51} & 98.94   & 25.87                   & \ding{51} & 46.71   & 17.69                   & \ding{51} & 92.18   & 24.76                   \\ \hline
            take a picture           & \ding{51} & 99.76   & 27.78             & \ding{51} & 99.84   & 27.78                   & \ding{51} & 69.12   & 23.11                   & \ding{51} & 60.07   & 23.96                   \\ \hline
            turn off the light       & \ding{51} & 76.09   & 24.81             & \ding{51} & 99.72   & 27.95                   & \ding{51} & 96.35   & 27.84                   & \ding{51} & 68.76   & 24.02                   \\ \hline
            turn on airplane mode    & \ding{51} & 75.66   & 15.61             & \ding{51} & 100.00  & 18.06                   & \ding{51} & 20.77   & 10.48                   & \ding{51} & 34.06   & 14.08                   \\ \Xhline{1px}
            Average                  & 10/10     & 89.85   & 23.59             & 10/10     & 99.28   & 25.34                   & 10/10     & 68.53   & 20.63                   & 10/10     & 63.14   & 21.08                   \\ \Xhline{1.5px}

            \multirow{2}{*}{Command} & \multicolumn{3}{c|}{Whisper (base)}     & \multicolumn{3}{c|}{Whisper (small)}          & \multicolumn{3}{c|}{Whisper (medium)}         & \multicolumn{3}{c}{Whisper (large)}           \\ \cline{2-13}
                                     & Attack    & TR (\%) & SNR (dB)          & Attack    & TR (\%) & SNR (dB)                & Attack    & TR (\%) & SNR (dB)                & Attack    & TR (\%) & SNR (dB)                \\ \Xhline{1px}
            call my wife             & \ding{51} & 58.90   & 13.55             & \ding{51} & 87.73   & 17.94                   & \ding{51} & 91.34   & 19.80                   & \ding{51} & 89.68   & 18.73                   \\ \hline
            make it warmer           & \ding{51} & 69.28   & 13.57             & \ding{51} & 78.24   & 15.00                   & \ding{51} & 80.67   & 18.76                   & \ding{51} & 81.56   & 17.89                   \\ \hline
            navigate to my home      & \ding{51} & 53.57   & 15.14             & \ding{51} & 77.49   & 19.11                   & \ding{51} & 85.16   & 22.64                   & \ding{51} & 91.71   & 25.44                   \\ \hline
            open the door            & \ding{51} & 72.91   & 18.62             & \ding{51} & 75.41   & 19.59                   & \ding{51} & 88.02   & 24.02                   & \ding{51} & 81.69   & 22.44                   \\ \hline
            open the website         & \ding{51} & 63.85   & 20.54             & \ding{51} & 78.41   & 22.53                   & \ding{51} & 89.61   & 25.15                   & \ding{51} & 94.93   & 26.76                   \\ \hline
            play music               & \ding{51} & 65.24   & 16.91             & \ding{51} & 81.73   & 23.04                   & \ding{51} & 88.06   & 26.81                   & \ding{51} & 91.75   & 26.85                   \\ \hline
            send a text              & \ding{51} & 55.08   & 19.94             & \ding{51} & 66.03   & 19.53                   & \ding{51} & 77.04   & 25.59                   & \ding{51} & 89.01   & 25.83                   \\ \hline
            take a picture           & \ding{51} & 84.77   & 26.71             & \ding{51} & 84.91   & 27.54                   & \ding{51} & 97.66   & 27.78                   & \ding{51} & 99.15   & 27.78                   \\ \hline
            turn off the light       & \ding{51} & 64.04   & 22.09             & \ding{51} & 86.75   & 25.97                   & \ding{51} & 95.12   & 26.45                   & \ding{51} & 97.20   & 27.61                   \\ \hline
            turn on airplane mode    & \ding{51} & 66.78   & 11.70             & \ding{51} & 92.72   & 17.32                   & \ding{51} & 96.96   & 16.92                   & \ding{51} & 99.18   & 18.06                   \\ \Xhline{1px}
            Average                  & 10/10     & 65.44   & 17.88             & 10/10     & 80.94   & 20.76                   & 10/10     & 88.97   & 23.39                   & 10/10     & 91.58   & 23.74                   \\ \Xhline{1.5px}
        \end{tabular}
    }
\end{table*}

In the evaluation, we observe a potential correlation between the performance of the target ASR system (\eg WER on LibriSpeech test-other) and the TR. 
To further analyze this correlation, we extend the set of 16 open-source ASRs to include Conformer-Transducer (XXL), Whisper (tiny), Whisper large-v2, and Whisper large-v3. Additionally, we categorize them into two groups: Whisper and Others. 
This categorization stems from the fact that Whisper is designed for multilingual recognition and has not undergone fine-tuning on the LibriSpeech dataset, unlike other open-source ASRs, which are tailored for English recognition tasks and include the LibriSpeech dataset as part of their training set.

\begin{figure}[!t]
    \centering
    \includegraphics[scale=0.4]{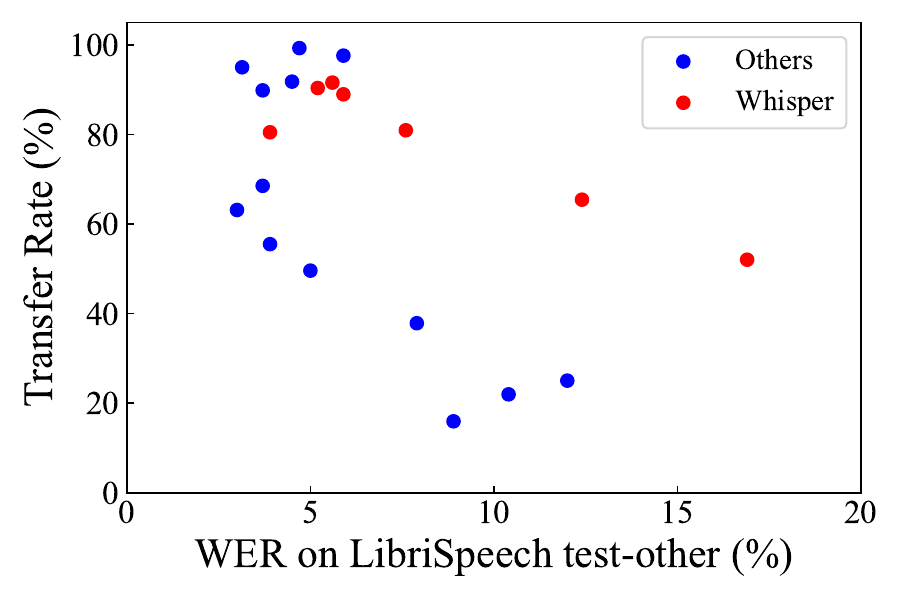}
    \caption{Correlation between ASR performance and TR.}
    \label{correlation_plot}
\end{figure}

The potential correlation is depicted in Figure~\ref{correlation_plot}, with Whisper and other open-source ASRs represented by different colors in the scatter plot.
It is discernible that there exists a potential negative correlation between the performance of the target ASR system and the TR.
For quantitative analysis, we employ the Pearson correlation coefficient \cite{cohen2009pearson} and Spearman’s rank correlation coefficient \cite{spearman1961proof} to characterize this correlation.
In statistics, the Pearson correlation coefficient measures the linear correlation between two sets of data, while Spearman’s rank correlation coefficient assesses the correlation of monotonic relationships.
Their values range from -1 to 1, with closer proximity to 1 indicating a stronger positive correlation, closer to -1 indicating a stronger negative correlation, and closer to 0 suggesting a weaker correlation.
The results are presented in Table~\ref{tab:pearson}.
They reveal a significant negative correlation between the performance and the TR for both categories of open-source ASRs.

We also conduct an ablation study to evaluate the impact of acoustic feature loss. Evaluations are carried out on 10 target commands, and 16 open-source ASR systems. There are 4 instances of attack failure (4/160) when the acoustic feature loss is omitted from the loss function.

\begin{table}[!t]
    \caption{Correlation coefficient between ASR performance and TR.}
    \label{tab:pearson}
    \centering
    \begin{tabular}{c|c|c}
        \Xhline{1px}
        Coefficient                 & Whisper & Others \\ \Xhline{1px}
        Pearson Correlation         & -0.933  & -0.758 \\ \hline
        Spearman's Rank Correlation & -0.607  & -0.580 \\ \Xhline{1px}
    \end{tabular}
\end{table}

\section{Adaptive search algorithm}

\begin{figure}[t!]
    \centering
    \includegraphics[scale=0.4]{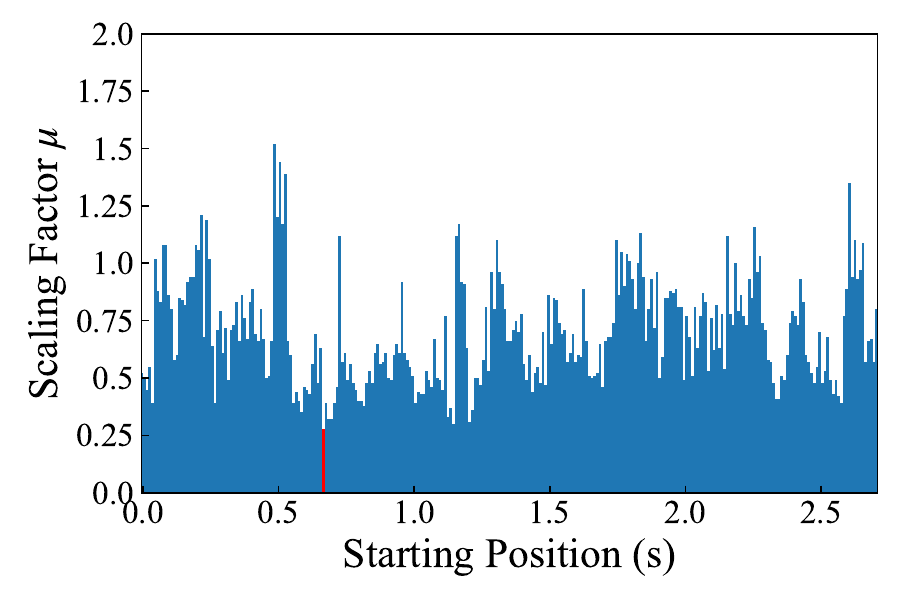}
    \caption{Scaling factor at different starting positions.}
    \label{delta_search}
\end{figure}

\begin{figure*}[t!]
    \centering
    \includegraphics[scale=0.4]{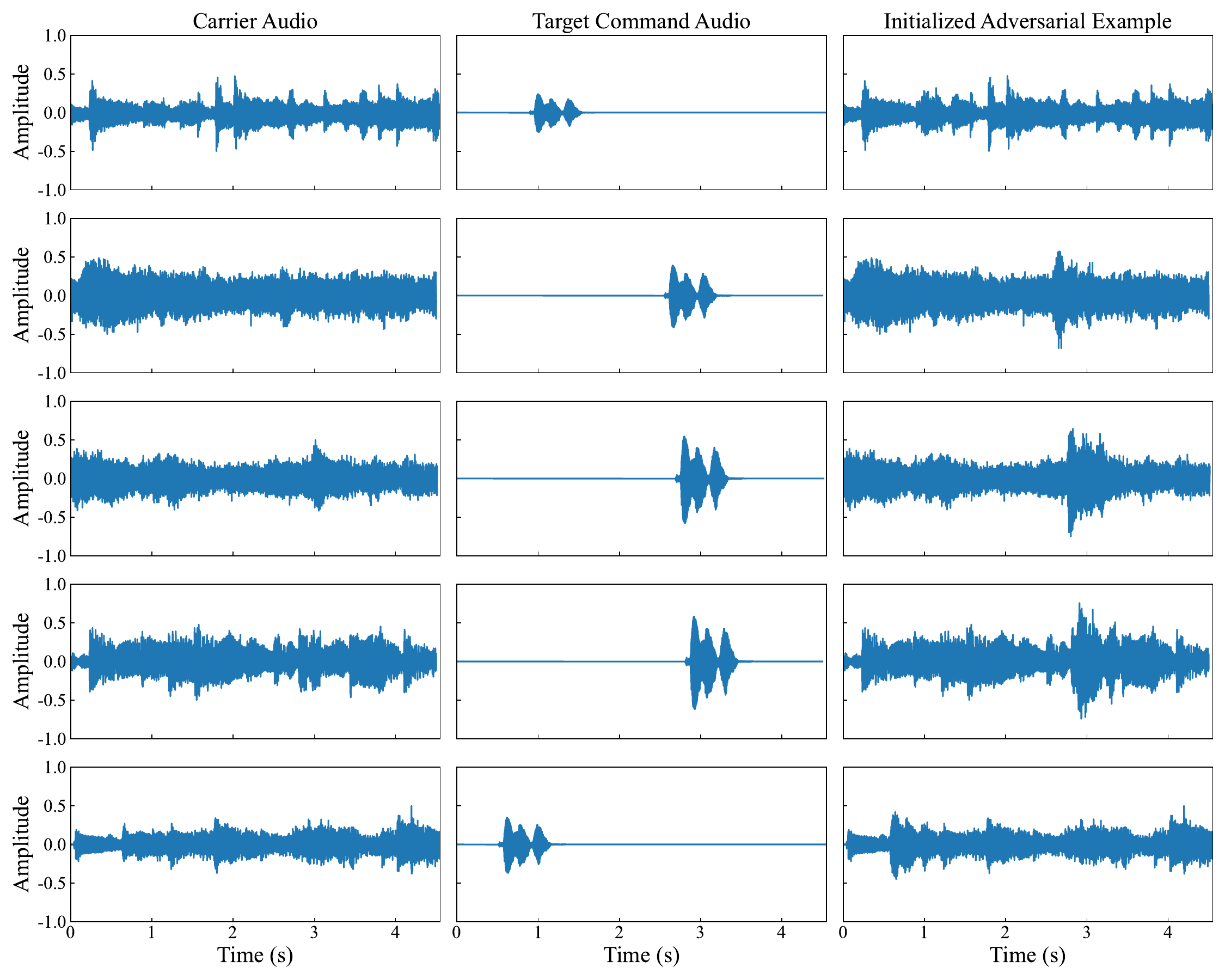}
    \caption{Waveforms of the initialized adversarial examples for different carrier audios.}
    \label{adaptive_initialization}
\end{figure*}

For a specific target command audio and carrier audio, we illustrate the values of scaling factor $\mu$ corresponding to different padding lengths in Figure~\ref{delta_search}. The starting position represents the length of the padding on the left side.
As the adaptive search algorithm is related to the carrier audio, we present the waveform of the initialized adversarial examples for the same target command audio under different carrier audios in Figure~\ref{adaptive_initialization}.
We can see that, for different carrier audios, the adaptive initialization algorithm finds distinct padding lengths on each side and scaling factors to initialize the adversarial perturbation.

Additionally, in our adaptive search algorithm, the obtained $\mu$ is a scalar. We also investigate treating $\mu$ as an optimizable variable.
Adversarial examples initialized using these two search methods and optimized through the sequential ensemble optimization algorithm exhibit comparable stealthiness and effectiveness.
However, treating $\mu$ as an optimizable variable results in greater time overhead. Therefore, we choose to search for $\mu$ as a scalar.

\end{document}